\documentclass[%
 amsmath,amssymb,reprint,twocolumn
]{revtex4}

\usepackage{subfigure}
\usepackage{epsfig}
\usepackage{exscale}
\usepackage{pbox}
\usepackage{multirow}
\usepackage{placeins}
\usepackage{graphicx}
\usepackage{dcolumn}
\usepackage{bm}
\newcolumntype{P}[1]{>{\centering\arraybackslash}p{#1}}
\newcolumntype{M}[1]{>{\centering\arraybackslash}m{#1}}

\begin{document}



\preprint{AIP/123-QED}

\title{Analysis and validation of low-frequency noise reduction in MOSFET circuits using variable duty cycle switched biasing}

\author{Kapil Jainwal}%
 \email{kapil.jainwal@lnmit.ac.in}
\author{Mukul Sarkar}%
 \email{msarkar@ee.iitd.ac.in}
\author{Kushal Shah}%
\email{kkshah@ee.iitd.ac.in}
\affiliation{Department of Electrical Engineering, Indian Institute of Technology (IIT) Delhi, Hauz Khas, New Delhi-110016, India}


\date{\today}

\begin{abstract}
Randomization of the trap state of defects present at the gate Si-SiO$_2$ interface of MOSFET is responsible for the  low-frequency noise phenomena such as Random Telegraph Signal (RTS), burst, and
1/\textit{f} noise. In a previous work, theoretical modelling and analysis of the RTS noise in MOS transistor was presented and it was shown that this 1/\textit{f} noise can be reduced by decreasing the
duty cycle ($f_{D}$) of switched biasing signal. In this paper, an extended analysis of this 1/\textit{f} noise reduction model is presented and it is shown that the RTS noise reduction is accompanied with shift in the corner frequency ($f_{c}$) of the 1/\textit{f} noise and the value of shift is a function of continuous ON time ({$T_{on}$}) of the device. This 1/\textit{f} noise reduction is also experimentally demonstrated in this paper using a circuit configuration with multiple identical transistor stages which produces a continuous output instead of a discrete signal. The circuit is
implemented in 180~nm standard CMOS technology, from UMC. According to the measurement results, the proposed technique reduces the 1/\textit{f} noise by approximately 5.9 dB at $f_{s}$ of 1~KHz for 2
stage, which is extended up to 16 dB at $f_{s}$ of 5 MHz for 6 stage configuration.

%
 \end{abstract}

 \pacs{Valid PACS appear here} 
\maketitle



\section{Introduction}
\label{sec:Intro}
The $1/f$ noise is a dominant noise source in the low-frequency region and is one of the major bottlenecks in applications like CMOS image sensors. The high noise limits the dynamic range of an image sensor. The primary sources of noise affecting an image sensor pixel are the thermal noise from the switches and the low-frequency $1/f$ noise from the in-pixel buffer. The thermal noise can be efficiently reduced using correlated double sampling (CDS) and an image sensor with 0.7e$^-$ noise has been reported \cite{paper:chen07erms}. The major limiting factor for the dynamic range now is the $1/f$ noise from the source follower (buffer). An image sensor is a time variant system due to time-varying biasing conditions. The low-frequency noise has an inverse relationship with frequency and aspect ratio  of the device, which becomes more prominent with decreasing dimensions of transistors  with technology scaling \cite{paper:Dunga08}. The decreasing dimensions of the transistors are needed to increase the spatial resolution of an image sensor and thus, reducing $1/f$ noise becomes more critical in improving the overall performance of a CMOS image sensor.

\par 
Random trapping and de-trapping of the charge carriers into the trap  defects present at the gate Si-SiO$_2$ interface of MOSFET is believed to be one of the major reasons responsible for the low-frequency noise phenomena such as random telegraph signal (RTS) and $1/f$ noise \cite{paper:lowfPRB, paper:ElectronicPtopertiesPRB, paper:TheoryexprmtfPRB, paper:keshner82, paper:Grasser12,paper:CHuScaling96}. There are a few models available which define the 1/$f$ noise conditionally but no model exists which can explain the $1/f$ noise phenomena completely \cite{paper:venderZeilBook86, paper:vanderzeril88, paper:Bulkeffectfnoise, paper:fcutoffparadoxPRB, paper:whatisNoise}.  Hooge's $ \Delta \mu $ model \cite{paper:Hog69} considers that the $1/f$ noise is caused by fluctuation in carrier mobility inside the bulk of MOS device. 
Carrier density fluctuation ($\Delta N$) model by McWhorter \cite{paper:McWh55} is based on the variation in the number of charge carriers inside the channel due to random behavior of the trap states present at the interface. This model states that the $1/f$ noise is a resultant of the RTS noise components from each trap. 
These traps are bias voltage dependent and have widely distributed emission/capture rates \cite{paper:Kirton892,paper:ManyElePRB}.  Due to time-varying biasing conditions, the non-stationarity is introduced in the behavior of the trap state, which makes the RTS noise and consequently the $1/f$ noise non-stationary. 
\par 
The $1/f$  noise in MOS transistors can be reduced by rapidly switching the device between accumulation and strong inversion region \cite{paper:Dierickx,paper:Venderwel07}. To model the noise for systems with time-varying biasing conditions, a proper stochastic model for trap state is required \cite{paper:Dierickx, paper:Venderwel07, paper:Bloom91,paper:ierkink99, paper:Klumperink2000, paper:VanderMEASUR2000,paper:OriginFPRB,paper:SHOTNOISEPRB,paper:MONTECARLOPRB}. H. Tian $et$  $al.$ \cite{paper:TianTHESIS} modeled the random activity of a single trap as a stochastic process and presented the first non-stationary RTS noise model using autocorrelation analysis of trap states in the time domain. The model in \cite{paper:TianTHESIS,paper:TianTCAS}, predicted that the noise reduction is independent of switching frequency ($f_s$) and concluded that the corner frequency ($f_c$) of the $1/f$ noise is independent of time-varying emission/capture rates. A. G. Mahmutoglu $et$ $at.$~\cite{paper:gockenNOISE} found that the limits taken in \cite{paper:TianTHESIS}, to calculate the time-averaged autocorrelation function (ACF) for trap state $N(t)$, were incorrect. In \cite{paper:gockenNOISE} and \cite{paper:Gocken14}, an improved analysis of the RTS noise was presented for switched biasing, using square wave (50\% duty cycle ($D$)) as input to a MOS transistor gate. In these and other papers on the low-frequency noise reduction \cite{paper:Dierickx, paper:Venderwel07, paper:Bloom91, paper:ierkink99, paper:Klumperink2000, paper:VanderMEASUR2000, paper:TianTHESIS, paper:TianTCAS, paper:gockenNOISE, paper:Gocken14, paper:klump_03, paper:reappr_conf, paper:JSSC_25percnt} the transistors are switched with either 50\%  or 25\% duty cycle and without multiple stages thereby leading to a discontinuous output. \par
If the device is kept ON for a shorter duration of time, it is obvious that the total noise will be less than what it would be if the device was always ON but this would lead to a discontinuous output.~What is required is a technique that can lead to noise reduction while keeping the output continuous. In~\cite{paper:Mypaper}, we have proposed to use variable duty cycle with multiple stages of transistors for continuous output and have shown that such configuration can lead to $1\big/f$ noise reduction. The $1/f$ noise is modeled for a complete range of duty cycle (0 to 100\%) of the switched biasing signal and its effect on the overall noise is studied.
Decreasing the duty cycle reduces the overall low-frequency noise by reducing the time for which the trap states are correlated. It is additionally shown that if multiple transistors are used with the same duty cycle (so as to ensure a continuous output), the total noise would still be less than what is obtained by using a single transistor which is ON all the time.  Thus, if $n$ transistors are ON for time $T\big/n$ (only one transistor is ON at any given time so as to ensure a continuous output), the total noise keeps decreasing as $n$ increases. This was shown using a mathematical model in our previous paper \cite{paper:Mypaper}. In this paper, an extended analysis of this $1/f$ noise model is presented and conclusions are verified with experimental results. A source follower implementation using multiple transistors, is presented to verify the noise reduction while maintaining a continuous time output.

\par 
The paper is organized as follows: A brief analysis of the RTS and $1/f$ noise is presented in Section \ref{sec:1/f}. The mathematical model for RTS noise model with variable duty cycle switched biasing is discussed in section \ref{sec:Trapnoise_model}. The $1/f$ noise reduction method using multiple transistors are described in section \ref{sec:fnoise_red}. In Section \ref{sec:CirDes} a circuit level implementation of the variable duty cycle switched biasing is presented using source follower transistor in standard CMOS image sensors. The simulation and experimental results are presented in section \ref{sec:Res}. The paper is concluded in section \ref{sec:conclusion}.

\begin{figure*} [ht!]
\centering

\def\putbox#1#2#3#4{\makebox[0in][l]{\makebox[#1][l]{}\raisebox{\baselineskip}[0in][0in]{\raisebox{#2}[0in][0in]{\scalebox{#3}{#4}}}}}
\def\rightbox#1{\makebox[0in][r]{#1}}
\def\centbox#1{\makebox[0in]{#1}}
\def\topbox#1{\raisebox{-0.60\baselineskip}[0in][0in]{#1}}
\def\midbox#1{\raisebox{-0.20\baselineskip}[0in][0in]{#1}}
\begin{center}
   \scalebox{0.453324}{
   \normalsize
   \parbox{14.3385in}{
   \includegraphics[scale=2.20593]{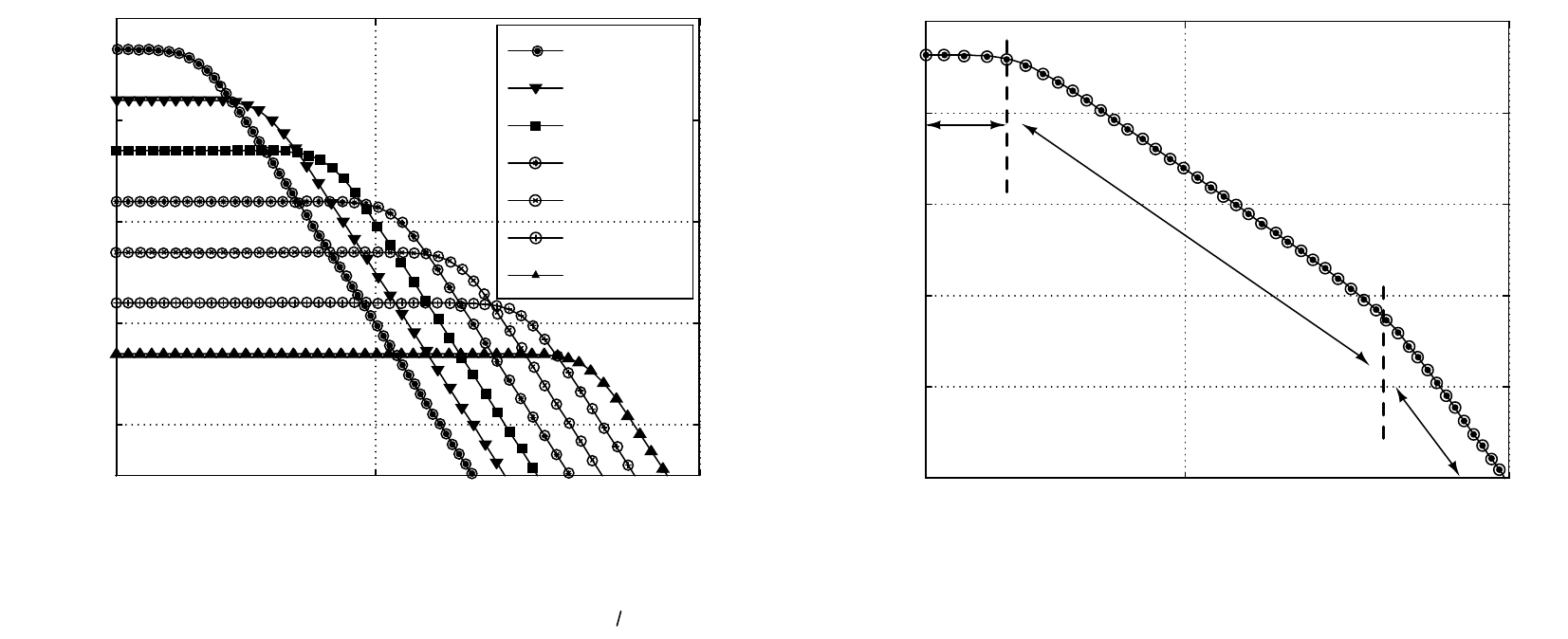}\\
   \putbox{-3.9  in}{1.2  in}{1.71}{\rotatebox{-360}{$10^5$}}%
   \putbox{-0.9  in}{1.2  in}{1.71}{\rotatebox{-360}{$10^{10}$}}%
   \putbox{-6.6  in}{1.88 in}{1.71}{\rotatebox{-360}{-100}}%
   \putbox{-6.56 in}{2.82 in}{1.71}{\rotatebox{-360}{-80}}%
   \putbox{-6.56 in}{3.77 in}{1.71}{\rotatebox{-360}{-60}}%
   \putbox{-6.56 in}{4.71 in}{1.71}{\rotatebox{-360}{-40}}%
   \putbox{-6.56 in}{5.66 in}{1.71}{\rotatebox{-360}{-20}}%
   \putbox{0.88 in}{1.38  in}{1.71}{-120}%
   \putbox{0.88 in}{2.23  in}{1.71}{-100}%
   \putbox{1.0  in}{3.08  in}{1.71}{-80}%
   \putbox{1.0  in}{3.93  in}{1.71}{-60}%
   \putbox{1.0  in}{4.78  in}{1.71}{-40}%
   \putbox{1.0  in}{5.63  in}{1.71}{-20}%
   \putbox{3.65 in}{1.2  in}{1.71}{$10^5$}%
   \putbox{6.7  in}{1.2  in}{1.71}{$10^{10}$}%
   \putbox{1.45 in}{4.41 in}{1.71}{(i) Flat}%
   \putbox{1.5 in}{4.11in}{1.71}{region}%
   \putbox{5.9  in}{3.28 in}{1.71}{(iii) $1/f^2$}%
   \putbox{6    in}{2.93in}{1.71}{region}%
   \putbox{2.6  in}{3.33 in}{1.71}{(ii) $1/f$ region}%
   \putbox{3.08 in}{0.71 in}{1.85}{Frequency [Hz]}%
   \putbox{-4.55 in}{0.75in}{1.85}{\rotatebox{-360}{Frequency [Hz]}}%
   \putbox{0.5 in}{2.45in}{1.85}{\rotatebox{-270}{$1/f$ Noise PSD [dB]}}%
   \putbox{-7  in}{2.33in}{1.85}{\rotatebox{-270}{RTS Noise PSD [dB]}}%
   \putbox{-6  in}{1.69in}{2.1}{\rotatebox{-360}{\textbf{(a)}}}%
   \putbox{1.5 in}{1.74in}{2.1}{\textbf{(b)}}%
   \putbox{-1.8 in}{5.39in}{1.71}{\rotatebox{-360}{$\lambda = 10^2$}}%
   \putbox{-1.8 in}{5.03in}{1.71}{\rotatebox{-360}{$\lambda = 10^3$}}%
   \putbox{-1.8 in}{4.69in}{1.71}{\rotatebox{-360}{$\lambda = 10^4$}}%
   \putbox{-1.8 in}{4.34in}{1.71}{\rotatebox{-360}{$\lambda = 10^5$}}%
   \putbox{-1.8 in}{3.99in}{1.71}{\rotatebox{-360}{$\lambda = 10^6$}}%
   \putbox{-1.8 in}{3.65in}{1.71}{\rotatebox{-360}{$\lambda = 10^7$}}%
   \putbox{-1.8 in}{3.30in}{1.71}{\rotatebox{-360}{$\lambda = 10^8$}}%
   } 
   } 
 \end{center}
  \vspace{- 0.4 in} 
  \caption{(a) The RTS noise PSD with different $\lambda$ or corner frequency ($f_c$). (b) The $1/f$ noise PSD (Both are according to the stationary noise model \cite{paper:McWh55}).}
     \label{fig:RTSnF}                                                                  
     \end{figure*}

\section{1/\textit{f} noise analysis, modeling and reduction technique}
\subsection{RTS and  1/\textit{f} noise analysis}
\label{sec:1/f} 
\par 
In order to understand the nature and mechanism of the RTS noise, the nature of the trap or defect state must be analyzed. Trapping and de-trapping of electrons into a single active trap can be characterized by an electron capture number or trap state $N(t)$, which is ``one'' when the trap is filled and  ``zero" otherwise. The probability of the trap being empty or filled is different when the transistor is ON or OFF. 

The trap capture rate $\lambda_c(t)$ and emission rate $\lambda_e(t)$ 
can be given as:
\begin{equation}
\lambda_c(t) = [1- P(t)]/\tau_c~;  \hspace{1cm}  \lambda_e(t) = P(t)/\tau_e,
\label{eq:rates}
\end{equation}
where $P(t)$ is the probability of trap occupancy (PTO) at time $t$. $\tau_c$ and $\tau_e$ are the  mean time before capture and mean time before emission of an electron, respectively \cite{paper:wangTHESIS}. $\tau_c$ is inversely proportional to the drain current. On the other hand $\tau_e$ is independent of the drain current. Due to the time-varying nature of the biasing condition and capture/emission meantime, the capture ($\lambda_c$) and the emission ($\lambda_e$) rates of the trap also become time variant \cite{paper:Kirton892}.  Note that in the above equation, the values of $\tau_e$ or $\tau_c$ are taken from the same distribution as for a stationary model and thus have no influence on the proposed noise reduction technique.
\par 
For a continuous ON  device with constant biasing, $\lambda_e = \lambda_c = \lambda$, which implies that probabilistically the trap would remain in each state (filled or empty) for approximately $\lambda^{-1}$ time. The value of ACF, for the samples separated by a time interval less than $\lambda^{-1}$ is high and increases as the time difference between samples decreases. This is due to the fact that the probability of the trap being in the same states is higher if the samples are taken within lesser time interval. If the samples are separated by time interval more than $\lambda^{-1}$, the correlation becomes weaker but never quite reaches zero. The ACF for $N(t)$ can be expressed as \cite{paper:venderZeilBook86}:

\begin{equation}
C_{\lambda} (\tau) = 0.25 e^{-2\lambda\tau}.
\label{eq:cstat}
\end{equation}

The double sided PSD corresponding to $C_{\lambda} (\tau)$ can be written as :
\begin{equation}
S_{\lambda} (f) = 0.25\lambda/(\lambda^2 + (\pi f)^2).
\label{eq:sstat}
\end{equation}
For stationary case the RTS noise spectrum is Lorentzian which is flat upto corner frequency ($\lambda$) and decays with a slope of -20 dB/decade for sampling frequencies above $\lambda$ (where $\lambda$ = ($\tau^{-1}_c$ + $\tau^{-1}_e$)/2$\pi$). 
The value of the stationary RTS noise can be given as \cite{paper:klump_03}:
\begin{equation}
S_{RTS} (f) = A^2_0.\frac{\beta}{(1+\beta)^2}.\frac{1}{4}.\frac{\lambda}{(\lambda^2 + (\pi f)^2)},
\end{equation}

where $A_0$ is RTS noise amplitude and $\beta=\tau_c/\tau_e$. 
The RTS noise plots for different values of $\lambda$ are shown in Fig.\ref{fig:RTSnF}(a). The RTS noise decreases for the sampling frequencies higher than $\lambda$ due to increase in correlation between samples with smaller time interval. Whereas, RTS noise PSD becomes flat for the sampling frequencies below $\lambda$ or corner frequency ($f_c$), due to very weak correlation between the sampled trap states.

\par The $1/f$ noise representation as a cumulative effect of the RTS noise from each trap is shown in Fig.~\ref{fig:RTSnF}(b). Multiple traps are present at the gate Si-SiO$_2$ interface with widely distributed trapping/de-trapping rates ranging from  $\lambda_{L}$ to $\lambda_{H}$. Due to these distributed rates, the $1/f$ noise plot can be divided into three different regions as shown in Fig. \ref{fig:RTSnF}(b).

 First (i) is the flat region, for the sampling frequencies below $\lambda_{L}$. The second region (ii) is between the frequency $\lambda_{L}$ and $\lambda_{H}$, where few traps are experiencing increasing correlation (reduction in the noise) with frequency while others have flat shaped RTS noise. In this region, the $1/f$ noise curve is proportional to the $f^{-1}$. In the final region (iii) (for the sampling frequencies above the $\lambda_{H}$), the Brownian noise is proportional to the $f^{-2}$. 


\subsection{Trap noise modeling for transistor with variable duty cycle switched biasing}
\label{sec:Trapnoise_model}

In~\cite{paper:gockenNOISE}, the RTS noise PSD is derived for a MOSFET device with switched biasing using square waveform (50~\% duty cycle) at the gate of the MOSFET as a control input. In \cite{paper:gockenNOISE} it is concluded that the RTS noise reduction depends on the $f_s$. In continuation with our previous work \cite{paper:Mypaper}, the RTS noise is modeled in a time-varying biasing condition for a more general case, where the duty cycle of switched bias signal varies between 0 and 100~\%. It is shown that the noise reduction depends on the continuous ON time ($T_{\text{\textit{on}}}$) of the device, which can be controlled by varying the duty cycle ($D$) of the periodically switched biasing signal. It is also concluded that the noise reduction is independent of $f_s$ for constant $T_{\text{\textit{on}}}$. The higher noise reduction can be achieved by decreasing the duty cycle for constant $f_s$, as compared to the model presented in \cite{paper:TianTHESIS,paper:gockenNOISE,paper:fnoisestationary1}.  As the variable biasing condition is periodic, the randomization of a single trap can be modeled by considering its behavior as a cyclo-stationary stochastic process \cite{paper:Enz07}. The capture and emission processes are random and governed by Poisson statistics. This Poisson process is inhomogeneous and exhibits non-stationarity, due to the dependency of $\lambda_c$ and $\lambda_e$ on voltage-dependent variables $\tau_e$ and $\tau_c$.        
\par 
For switched biasing, $\lambda_{e}=\lambda_{c}=\lambda_{\text{\textit{on}}}$ for ON state and $\lambda_{e}= \lambda_{\text{\textit{off}}}$, $\lambda_{c} \approx 0$ for OFF state of the device. The value of ACF is almost zero during the OFF state of the device. Switching the transistor OFF between consecutive ON time period, for sufficient time ($\approx \lambda^{-1}_{\text{\textit{off}}}$), resets the probability of trap occupancy to zero. The value of the PTO is close to  zero (during OFF state of the device) due to very high emission rate in the absence of conduction. If the periodic ON time ($T_{\text{\textit{on}}}$) is less than $1/\lambda_{\text{\textit{on}}}$, then the trap state are reset in every $T_{rst} = $ $T_{\text{\textit{on}}}$ + $\lambda^{-1}_{\text{\textit{off}}}$ $\approx$ $T_{\text{\textit{on}}}$ (for $\lambda^{-1}_{\text{\textit{off}}}$ $<<$ $T_{\text{\textit{on}}}$). Thus, the trap states separated by $T_{\text{\textit{rst}}}$ $\approx$ $T_{\text{\textit{on}}}$ time interval or more, have zero correlation. 

\begin{figure}
\centering
\def\putbox#1#2#3#4{\makebox[0in][l]{\makebox[#1][l]{}\raisebox{\baselineskip}[0in][0in]{\raisebox{#2}[0in][0in]{\scalebox{#3}{#4}}}}}
\def\rightbox#1{\makebox[0in][r]{#1}}
\def\centbox#1{\makebox[0in]{#1}}
\def\topbox#1{\raisebox{-0.60\baselineskip}[0in][0in]{#1}}
\def\midbox#1{\raisebox{-0.20\baselineskip}[0in][0in]{#1}}

   \scalebox{0.4511}{
   \footnotesize
   \parbox{7.45091111111111 in}{
   \includegraphics[scale=2.2168]{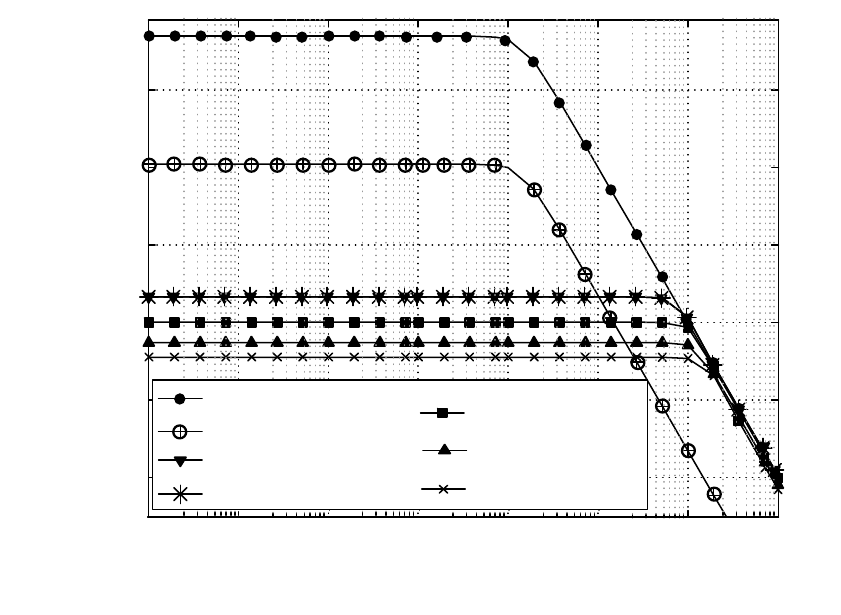}\\
   \putbox{-2.5 in}{2.59in}{2.2168}{\rightbox{\midbox{-160}}}%
   \putbox{-2.5 in}{3.29in}{2.2168}{\rightbox{\midbox{-140}}}%
   \putbox{-2.5 in}{3.97in}{2.2168}{\rightbox{\midbox{-120}}}%
   \putbox{-2.5 in}{4.65in}{2.2168}{\rightbox{\midbox{-100}}}%
   \putbox{-2.5 in}{5.27in}{2.2168}{\rightbox{\midbox{-80}}}%
   \putbox{-3.4 in}{3.17in}{2.66016}{\rotatebox{-270}{\centbox{RTS Noise PSD [dB]}}}%
   \putbox{-2.43 in}{0.76in}{2.2168}{\centbox{\topbox{$10^{0}$}}}%
   \putbox{-0.25 in}{0.25in}{2.66016}{\midbox{Frequency [Hz]}}%
   \putbox{-1.58 in}{0.75in}{2.2168}{\centbox{\topbox{$10^{2}$}}}%
   \putbox{1.61 in}{0.74in}{2.2168}{\centbox{\topbox{$10^{10}$}}}%
   \putbox{2.41 in}{0.74in}{2.2168}{\centbox{\topbox{$10^{12}$}}}%
   \putbox{3.2 in}{0.74in}{2.2168}{\centbox{\topbox{$10^{14}$}}}%
   \putbox{-2.5 in}{1.90in}{2.2168}{\rightbox{\midbox{-180}}}%
   \putbox{-2.5 in}{1.21in}{2.2168}{\rightbox{\midbox{-200}}}%
   \putbox{-1.8 in}{1.91in}{2.}{\midbox{Stationary PSD}}%
   \putbox{-1.85 in}{1.60in}{2.}{\midbox{~\cite{paper:TianTHESIS} ($D$ = 50~\%)}}%
   \putbox{-1.85 in}{1.34in}{2.}{\midbox{~\cite{paper:gockenNOISE} ($D$ = 50~\%)}}%
   \putbox{-1.8 in}{1.05in}{2.}{\midbox{$D$ = 50~\% }}%
   \putbox{0.5 in}{1.75in}{2.}{\midbox{$D$ = 25~\% }}%
   \putbox{0.5 in}{1.42in}{2.}{\midbox{$D$ = 12.5~\% }}%
   \putbox{0.5 in}{1.10in}{2.}{\midbox{$D$ = 6.25~\% }}%
   \putbox{-0.78 in}{0.75in}{2.2168}{\centbox{\topbox{$10^{4}$}}}%
   \putbox{0 in}{0.75in}{2.2168}{\centbox{\topbox{$10^{6}$}}}%
   \putbox{0.81 in}{0.75in}{2.2168}{\centbox{\topbox{$10^{8}$}}}%
   } 
   } 
  \caption{The RTS noise trapped spectrum $S^{s}_{\lambda}(\omega)$ evaluated from Eq. (\ref{eq:Ssw}) [MATLAB simulation]: For single transistor with constant (DC) and switched biasing with variable duty cycle ($D$)} .
\label{fig:RTS} 
\vspace{-0.1 in} 
\end{figure}

For sampling frequencies below $T^{-1}_{\text{\textit{on}}}$, the  region in the noise spectrum can be seen in Fig. \ref{fig:RTS}, and above that the trap states are correlated and the switched noise PSD curve overlaps with stationary noise PSD curve~\cite{paper:fnoisestationary1, paper:fstationary2}. The reduction in the noise occurs by  
varying correlation due to which the noise PSD becomes flat for the sampling frequencies below $T^{-1}_{\text{\textit{on}}}$ Hz. The corner frequency of the noise is thus limited by $T_{\text{\textit{on}}}$. From the above analysis, it can be stated that the noise reduction increases with a decrease in value of $T_{\text{\textit{on}}}$ or duty cycle of the switched biasing signal and following conclusions can be derived.
%
\begin{enumerate}
\item Decrease in the $1/f$ noise power sampling frequencies above $T^{-1}_{\text{\textit{on}}}$ is due to the increasing correlation between trap states.  
Due to switching action, the correlation between the samples, separated by more than $T_{on}$ time, becomes very weak. This makes the noise PSD ``flat" for the sampling frequencies $T^{-1}_{on}$. Hence, The corner frequency is shifted to a new value of $T^{-1}_{\text{\textit{on}}}$ for $T_{\text{\textit{on}}} < \lambda^{-1}_{\text{\textit{on}}}$, which causes reduction in the noise power. 

\item For the trap states of a single transistor sampled during ON time and separated by time interval less than T$ _{on} $, have the same correlation as DC biasing condition. Hence, the noise power for sampling frequencies above the $T^{-1}_{\text{\textit{on}}}$ is 10log($n$)~dB less than  the noise power of the stationary noise PSD (reduction of 10log($n$) dB in the noise power is due to OFF state of the device).

\item A higher noise reduction can be achieved by lowering the value of $T_{\text{\textit{on}}}$ either by increasing $f_s$ or decreasing duty cycle. If $T_{on}$ is kept constant by suitably varying the duty cycle the obtained noise reduction is also constant for varying $f_s$. 

\end{enumerate}

The RTS noise modeling is given as follows:
\begin{itemize}
\item The probability of trap being filled in ON state is:
\end{itemize}
\begin{equation}
P_{\text{\textit{on}}}(t)= 0.5 + a e^{-2\lambda_{\text{\textit{on}}}t}, \hspace{0.5cm} \text{for}~\hspace{0.3cm} 0 \leq t < T/n,
\label{eq:pon}
\end{equation}
where `$a$' is a constant which is dependent on the PTO at initial condition ($t = 0$) of the ON state, `$T$' is the time period of the biasing signal, `$n$' is an integer given by $n$ = $T/T_{\text{\textit{on}}}$.
\begin{itemize}
 \item The probability of trap to be filled in OFF state is:
\end{itemize}
\begin{equation}
P_{\text{\textit{off}}}(t)= b e^{-\lambda_{\text{\textit{off}}}t} \hspace{0.5cm} \text{for}~\hspace{0.3cm}  T/n \leq t < T, \vspace{- 0.05 in}
\label{eq:poff}
\end{equation}
where `$b$' is a constant which is dependent on the PTO at initial condition of the OFF state. 

The value of initial conditions of `$a$' and `$b$' can be derived as:
\begin{equation}
\begin{aligned}
a = -(1-\beta)/(1-\alpha\beta); \hspace{0.3cm}  b = (1-\alpha)/(1-\alpha\beta),
\label{eq:ab}
\end{aligned}
\end{equation}
\begin{equation}
\text{where}~~ \alpha = e^{ -2\lambda_{\text{\textit{on}}} \frac{T}{n} }; \hspace{0.3cm} \beta = e^{ -\lambda_{\text{\textit{off}}} \frac{(n-1)T}{n}. } \nonumber \hspace{2.5cm} \vspace{- 0.05 in}
\end{equation}
\par
In the usual cyclo-stationary noise models only 50\% duty cycle and ON time noise PSD is considered. In the OFF state, the traps remain empty due to very high emission rate and negligibly small capture rate. This makes the initial probability of trap occupancy (PTO) at the the beginning of ON time to be almost zero, if $\lambda_{\text{\textit{off}}}T_{\text{\textit{off}}}~>>~1$. In the proposed model, the ON time and OFF time may not be equal, as the OFF time of the device ($T_{\text{\textit{off}}}$) increases as duty cycle decreases. The PTO before commencement of ON time becomes negligibly small ($P_{on}(0)~\approx~0$). Thus for ON time PSD the value of variable `a' would be -0.5. The PTO at the initial condition of OFF state is equal to the PTO at time $T/n$ during ON state which can be calculated by Eq. (\ref{eq:pon}). The noise PSD during OFF state can be calculated with this initial condition. The PTO decreases exponentially and can be given by Eq. (\ref{eq:poff}). As $\lambda_{\text{\textit{off}}}$ is very high the PTO at time equal to $T(n-1)/n$ during OFF state, is negligibly small (considered as zero). Thus, OFF time PSD can be considered as negligible.

From Eqs.~(\ref{eq:pon}), (\ref{eq:poff}), and (\ref{eq:ab}), it can be seen that the PTO function depends on capture/emission rates, switching frequency ($1/T $) and initial condition of the trap state. Using these equations the values of `$a$' and `$b$' obtained are -$0.5$ and $0$ respectively, with the assumption that $\lambda_{\text{\textit{on}}}T_{\text{\textit{on}}} << 1$ and
$\lambda_{\text{\textit{off}}}T_{\text{\textit{off}}} >> 1$ as in~\cite{paper:Menglia04, paper:zanolla10}.

Trap state $N(t)$ is a cyclo-stationary random process, whose ACF is given by~\cite{paper:papoulis}.
During ON state:
\begin{equation}
\text{For} \hspace{0.2cm} \left|\tau/{2}\right|  \leq t < T/n - \left|\tau/2\right|, \hspace{5cm} \nonumber
\label{eq:limits}
\end{equation}
\begin{equation}
C_{\lambda_{\text{\textit{on}}}}(t,\tau) = (1/4) e^{-2\lambda_{\text{\textit{on}}}\left|\tau\right|} - a^2 e^{-4\lambda_{\text{\textit{on}}}t}.  \hspace{1.5 cm}
\label{eq:con(lt)}
\end{equation}
The double sided PSD of the RTS noise of a single trap is calculated using the Fourier transform of the ACF during the ON time period of the transistor.
\begin{equation}
S_{\lambda,{on}}(\omega)= \mathcal{F}[{C^{s}_{\text{\textit{on}}}(\tau)}],
\label{eq:Sw}
\end{equation}
where $C^{s}_{\text{\textit{on}}}(\tau)$ is the time average values of $C_{\lambda_{\text{\textit{on}}}}(t,\tau)$.
\begin{equation}
\begin{aligned}
S_{\lambda,{on}}(\omega) =
\frac{1/4T}{(4\lambda^2_{\text{\textit{on}}}+\omega^2)} \Biggl[ \frac{4T\lambda_{\text{\textit{on}}}}{n}-A -\frac{Be^{\frac{-2\lambda_{\text{\textit{on}}}T}{n}}}{4 \lambda^2_{\text{\textit{on}}}+\omega^2} \Biggr],
\end{aligned}
\label{eq:Son}
\end{equation}
\begin{equation}
A = a^2 4e^{\frac{-4\lambda_{\text{\textit{on}}}T}{n}} + \frac{ (16a^2 + 8)\lambda^2_{\text{\textit{on}}} + a^24\omega^2 -2\omega^2}{(4\lambda^2_{\text{\textit{on}}}+\omega^2)}, \nonumber
\end{equation}
\begin{equation}
B = (40a^2 + 6) \lambda^2_{\text{\textit{on}}}cos\Big(\frac{\omega T}{n}\Big) - 8\omega \lambda_{\text{\textit{on}}}sin\Big(\frac{\omega T}{n}\Big). \nonumber
\end{equation}
In a similar way the noise PSD $S_{\lambda,{\text{\textit{off}}}}(\omega)$, for OFF state, can also be calculated by taking Fourier transformation of time average of $C_{\lambda_{\text{\textit{off}}}}(t,\tau)$. In this work, a variable duty cycle is used for biasing signal in which OFF time is higher. As OFF time increases the probability of trap to be empty is high due to higher $\lambda_{\text{\textit{off}}}$ and to get filled again is very low as $\lambda_{\text{\textit{on}}}$ is very low \cite{paper:Menglia04,paper:zanolla10}. Thus, the noise in OFF state of the device can be neglected and the RTS noise PSD can be given as: 

\begin{equation}
S^{s}_{\lambda}(\omega)~\approx~S_{\lambda,{on}}(\omega).
\label{eq:Ssw}
\end{equation}

According to Eq. (\ref{eq:Son}) the RTS noise PSD of a single trap is a function of the duty cycle of the switched bias signal ($T/n$), emission/capture rate, and initial condition of PTO during ON time.

\par
The $1/f$ noise is calculated by superposition of the noise generated by individual traps with different capture/emission rates. The overall $1/f$ noise PSD is given by \cite{paper:Kirton892}:
\begin{equation}
S(\omega) = \int_{\lambda_{L}}^{\lambda_{H}} S^{s}_{\lambda}(\omega)g(\lambda)d\lambda.
\label{eq:Sformula}
\end{equation}

\begin{equation}
g(\lambda) = \frac{4k\theta At_{ox} N_t}{\lambda\log( \frac {\lambda_{H}}{\lambda_{L}})},
\label{eq:gformula}
\end{equation}
where $g(\lambda)$ is the distribution function of the emission and capture rate, $\theta$ is the absolute temperature in Kelvin, $k$ is Boltzmann constant, $A$ is the channel area (1 $\mu m^2$), $t_{ox}$ is the effective gate oxide thickness ($10~nm$), $N_t$ is the trap density, $\lambda_H$ and $\lambda_L$ are the fastest and slowest transition rates, respectively. These rates are related to $t_{ox}$ through the equation log($\lambda_H$/$\lambda_L$) = $\gamma$$t_{ox}$, where $\gamma$ is the tunneling constant.
The $1/f$ noise voltage PSD \cite{paper:TianTHESIS} is given by:
\begin{equation}
S_{1/f} (\omega) = \Big(\frac{q}{AC_{ox}}\Big)^2 S(\omega),
\label{eq:s1overf}
\end{equation}
where $C_{ox}$ is the unit area channel capacitance and $q$ is electron charge.

Equations (\ref{eq:Son}), (\ref{eq:Ssw}), and (\ref{eq:s1overf}) are valid for the entire range of sampling frequency including the frequencies below and above the switching frequency. The entire range is selected to verify if the noise reduction due to duty cycling with multiple stages happens at all frequencies or only at specific frequencies. In our analysis, the noise reduction is observed only for frequencies below the corner frequency. Above the corner frequency, the noise remains 10log($n$) dB less than the stationary noise due to OFF state.

\subsection{ 1/\textit{f} noise reduction by using multiple transistors with varying duty cycle switched biasing}
\label{sec:fnoise_red}
As discussed in section \ref{sec:Trapnoise_model}, the $1/f$ noise PSD decreases with a decrease in the duty cycle ($D$) of the switched biasing signal. The derived model is also validated as higher RTS noise reduction can be obtained with an increase in the values of `$n$', as per (11). Multiple transistors are connected between the input and the output wherein the noise contribution of each transistor is assumed to be non-correlated. Thus overall $1/f$ noise PSDs would be the summation of the noise PSDs of each source follower and is given as:
\begin{equation}
S^{o/p}_{\lambda} (\omega) = \textit{n} \times S_{1/f} (\omega),
\label{eq:Sn}
\end{equation}
where $S_{1/f} (\omega)$ is the worst case noise PSD of each stage.
\par
In \cite{paper:wangTHESIS}, $\lambda_{\text{\textit{on}}}$ = $10^8$~Hz,~$\lambda_{\text{\textit{off}}}=10^{-20}$~Hz, and $T$ = $10^{-11}$ $s$ (with 50 \% duty cycle) is selected such that $\lambda_{\text{\textit{off}}}T_{\text{\textit{off}}}$ $>>$ 1 and $\lambda_{\text{\textit{on}}}T_{\text{\textit{on}}}$ $<<$ 1. The reason behind these limitations can be explained from the analysis shown in section \ref{sec:Trapnoise_model}. If $\lambda_{\text{\textit{on}}}T_{\text{\textit{on}}}$ $<<$ 1, trap states are getting reset by switching. However, $\lambda_{\text{\textit{off}}}$ and $\lambda_{\text{\textit{on}}}$ depend on time, biasing voltage and are different for each trap. $\lambda_{\text{\textit{off}}}T_{\text{\textit{off}}}$ $>>$ 1 is true only when $T_{\text{\textit{off}}}$ and $\lambda_{\text{\textit{off}}}$ are very large. If this condition does not hold, then there may be a non-zero probability of trap to remain filled until the starting of the next ON time of the transistor. If the initial PTO for ON state of the device is not zero, the noise PSD does not remain the same as in \cite{paper:gockenNOISE}.
 
\begin{figure*}[ht!]
\centering

\def\putbox#1#2#3#4{\makebox[0in][l]{\makebox[#1][l]{}\raisebox{\baselineskip}[0in][0in]{\raisebox{#2}[0in][0in]{\scalebox{#3}{#4}}}}}
\def\rightbox#1{\makebox[0in][r]{#1}}
\def\centbox#1{\makebox[0in]{#1}}
\def\topbox#1{\raisebox{-0.60\baselineskip}[0in][0in]{#1}}
\def\midbox#1{\raisebox{-0.20\baselineskip}[0in][0in]{#1}}
   \scalebox{0.702929}{
   \normalsize
   \parbox{9.95833in}{
   \includegraphics[scale=1.32]{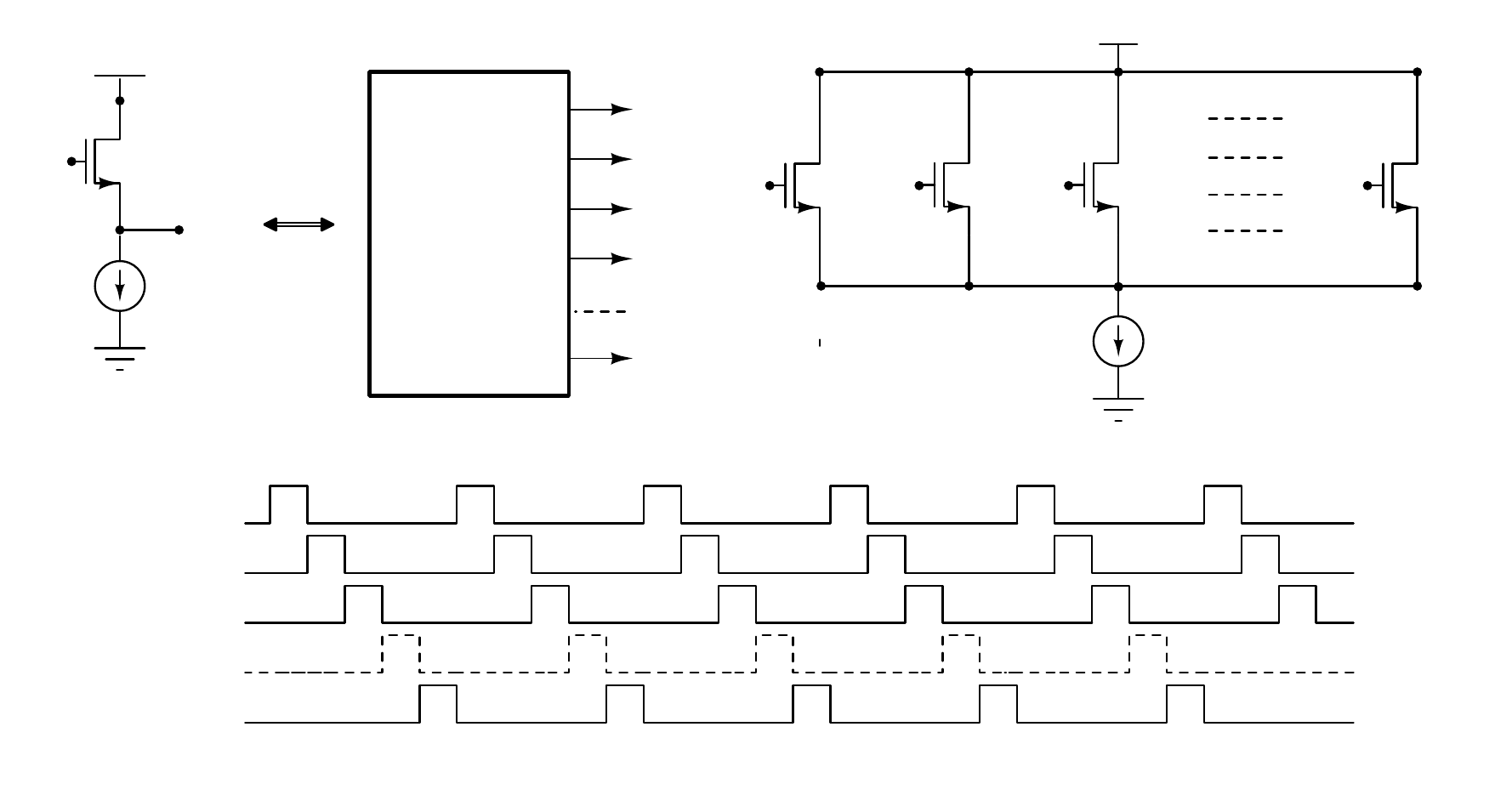}\\
   \putbox{-4.66 in}{4.0in}{1.56}{\midbox{$V_{in}$}}%
   \putbox{-4.15 in}{4.6 in}{1.56}{\midbox{$V_{DD}$}}%
   \putbox{-4.05 in}{2.4 in}{1.50}{\midbox{(a)}}%
   \putbox{-3.7 in}{3.15 in}{1.56}{\midbox{$I_{dc}$}}%
   \putbox{-3.5 in}{3.5 in}{1.56}{\midbox{$V_{out}$}}%
   \putbox{2.5 in}{2.8 in}{1.56}{\midbox{$I_{dc}$}}%
   \putbox{2.3 in}{3.3 in}{1.56}{\midbox{$V'_{out}$}}%
   \putbox{2.05 in}{4.8 in}{1.56}{\midbox{$V_{DD}$}}%
   \putbox{-1.1 in}{4.4 in}{1.56}{\midbox{$V_{clk,1}$}}%
   \putbox{-1.1 in}{4.10 in}{1.56}{\midbox{$V_{clk,2}$}}%
   \putbox{-1.1 in}{3.8 in}{1.56}{\midbox{$V_{clk,3}$}}%
   \putbox{-1.1 in}{3.5 in}{1.56}{\midbox{$V_{clk,4}$}}%
   \putbox{-1.1 in}{2.85 in}{1.56}{\midbox{$V_{clk,n}$}}%
   \putbox{-0.4 in}{3.7 in}{1.56}{\midbox{$V_{clk,1}$}}%
   \putbox{0.5 in}{3.7 in}{1.56}{\midbox{$V_{clk,2}$}}%
   \putbox{1.4 in}{3.7 in}{1.56}{\midbox{$V_{clk,3}$}}%
   \putbox{3.3 in}{3.7 in}{1.56}{\midbox{$V_{clk,n}$}}%
   \putbox{-3.8 in}{1.85in}{1.80}{\midbox{$V_{clk,1}$}}%
   \putbox{-3.8 in}{1.52in}{1.80}{\midbox{$V_{clk,2}$}}%
   \putbox{-3.8 in}{1.18in}{1.80}{\midbox{$V_{clk,3}$}}%
   \putbox{-3.8 in}{0.51in}{1.80}{\midbox{$V_{clk,n}$}}%
   \putbox{2.1 in}{2.1 in}{1.50}{\midbox{(b)}}%
   \putbox{0.29 in}{0.22in}{1.50}{\midbox{(c)}}%
   \putbox{-2.35 in}{4.09in}{1.35}{\midbox{Programmable}}%
   \putbox{-1.95 in}{3.78in}{1.35}{\midbox{Ring}}%
   \putbox{-2.05 in}{3.49in}{1.35}{\midbox{Counter}}%
   } 
   } 
\caption{(a) Single transistor source follower, (b) SF action achieved through multiple transistors with programmable ring counter, and (c) Timing diagram. }
\label{fig:SF}                                                           
\end{figure*}

\begin{figure*}[ht!]
\centering

\def\putbox#1#2#3#4{\makebox[0in][l]{\makebox[#1][l]{}\raisebox{\baselineskip}[0in][0in]{\raisebox{#2}[0in][0in]{\scalebox{#3}{#4}}}}}
\def\rightbox#1{\makebox[0in][r]{#1}}
\def\centbox#1{\makebox[0in]{#1}}
\def\topbox#1{\raisebox{-0.60\baselineskip}[0in][0in]{#1}}
\def\midbox#1{\raisebox{-0.20\baselineskip}[0in][0in]{#1}}
   \scalebox{0.3426}{
   \normalsize
   \parbox{22.276in}{
   \includegraphics[scale=2.91886]{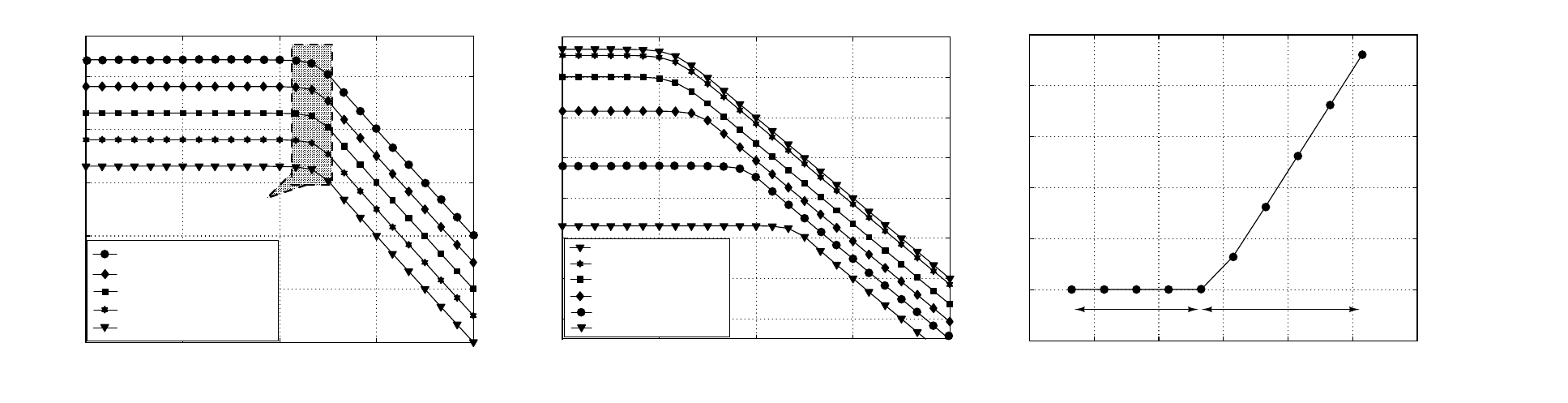}\\
   \putbox{-10.08 in}{0.55in}{1.68}{$10^6$}%
   \putbox{-8.68 in}{0.55in}{1.68}{$10^8$}%
   \putbox{-7.32 in}{0.55in}{1.68}{$10^{10}$}%
   \putbox{-5.92 in}{0.55in}{1.68}{$10^{12}$}%
   \putbox{-4.52 in}{0.55in}{1.68}{$10^{14}$}%
   \putbox{-10.5 in}{0.81in}{1.68}{-240}%
   \putbox{-10.5 in}{1.57in}{1.68}{-220}%
   \putbox{-10.5 in}{2.34in}{1.68}{-200}%
   \putbox{-10.5 in}{3.10in}{1.68}{-180}%
   \putbox{-10.5 in}{3.87in}{1.68}{-160}%
   \putbox{-10.5 in}{4.64in}{1.68}{-140}%
   \putbox{-3.21 in}{0.62in}{1.68}{$10^{6}$}%
   \putbox{-1.82 in}{0.62in}{1.68}{$10^{8}$}%
   \putbox{-0.45 in}{0.62in}{1.68}{$10^{10}$}%
   \putbox{0.95 in}{0.62in}{1.68}{$10^{12}$}%
   \putbox{2.34 in}{0.62in}{1.68}{$10^{14}$}%
   \putbox{-3.63 in}{1.14in}{1.68}{-220}%
   \putbox{-3.63 in}{1.72in}{1.68}{-200}%
   \putbox{-3.63 in}{2.30in}{1.68}{-180}%
   \putbox{-3.63 in}{2.88in}{1.68}{-160}%
   \putbox{-3.63 in}{3.46in}{1.68}{-140}%
   \putbox{-3.63 in}{4.04in}{1.68}{-120}%
   \putbox{-3.63 in}{4.62in}{1.68}{-100}%
   \putbox{-3.63 in}{5.20in}{1.44}{-80}%
   \putbox{-1.18 in}{0.24in}{1.92}{Frequency [Hz]}%
   \putbox{3.6 in}{0.60in}{1.68}{2}%
   \putbox{4.53 in}{0.60in}{1.68}{4}%
   \putbox{5.46 in}{0.60in}{1.68}{6}%
   \putbox{6.39 in}{0.60in}{1.68}{8}%
   \putbox{7.27 in}{0.60in}{1.68}{10}%
   \putbox{8.21 in}{0.60in}{1.68}{12}%
   \putbox{9.12 in}{0.60in}{1.68}{14}%
   \putbox{3.41 in}{0.81in}{1.68}{7}%
   \putbox{3.41 in}{1.54in}{1.68}{8}%
   \putbox{3.41 in}{2.27in}{1.68}{9}%
   \putbox{3.32 in}{3.01in}{1.68}{10}%
   \putbox{3.32 in}{3.74in}{1.68}{11}%
   \putbox{3.32 in}{4.47in}{1.68}{12}%
   \putbox{3.32 in}{5.21in}{1.68}{13}%
   \putbox{-8.08 in}{0.21in}{1.92}{Frequency [Hz]}%
   \putbox{-9.43 in}{2.09in}{1.44}{$T$ = $10^{-6}~s$, $D$ = 0.005$~\%$}%
   \putbox{-9.43 in}{1.81in}{1.44}{$T$ = $10^{-7}~s$, $D$ = 0.05$~\%$}%
   \putbox{-9.43 in}{1.55in}{1.44}{$T$ = $10^{-8}~s$, $D$ = 0.5$~\%$}%
   \putbox{-9.43 in}{1.29in}{1.44}{$T$ = $10^{-9}~s$, $D$ = 5$~\%$}%
   \putbox{-9.43 in}{1.02in}{1.44}{$T$ = $10^{-10}~s$, $D$ = 50$~\%$}%
   \putbox{-9.92 in}{5.43in}{1.68}{$T_{\text{\textit{on}}}$ = 5$\times10^{-11}~s$, $\lambda_{\text{\textit{on}}}$ = $10^8$~Hz  }%
   \putbox{-9.61 in}{2.95in}{1.68}{Equal shift in $f_c$ for   }%
   \putbox{-9.98 in}{2.66in}{1.68}{Equal $T_{\text{\textit{on}}}$ and Different $T$ (or $f_s$)  }%
   \putbox{-2.94 in}{5.47in}{1.68}{$T$ =  $10^{-5}~s$, $T_{\text{\textit{on}}}$ is variable, and $\lambda_{\text{\textit{on}}}$ = $10^8$~Hz  }%
   \putbox{-2.59 in}{2.17in}{1.44}{Stationary Noise PSD}%
   \putbox{-2.59 in}{1.94in}{1.44}{$T_{\text{\textit{on}}}$ = 5$\times10^{-6}~s$}%
   \putbox{-2.59 in}{1.71in}{1.44}{$T_{\text{\textit{on}}}$ = 5$\times10^{-7}~s$}%
   \putbox{-2.59 in}{1.48in}{1.44}{$T_{\text{\textit{on}}}$ = 5$\times10^{-8}~s$}%
   \putbox{-2.59 in}{1.02in}{1.44}{$T_{\text{\textit{on}}}$ = 5$\times10^{-10}~s$}%
   \putbox{-2.59 in}{1.25in}{1.44}{$T_{\text{\textit{on}}}$ = 5$\times10^{-9}~s$}%
   \putbox{5.77 in}{0.24in}{1.68}{$-log(T_{\text{\textit{on}}})$}%
   \putbox{2.92 in}{2.62in}{1.68}{\rotatebox{-270}{$log(f_c)$}}%
   \putbox{-3.89 in}{2.15in}{1.68}{\rotatebox{-270}{RTS Noise PSD [dB]}}%
   \putbox{-10.83 in}{2.07in}{1.68}{\rotatebox{-270}{RTS Noise PSD [dB]}}%
   \putbox{-4.87 in}{4.98in}{2.5}{\textbf{(a)}}%
   \putbox{2.01 in}{4.95in}{2.5}{\textbf{(b)}}%
   \putbox{8.6 in}{4.95in}{2.5}{\textbf{(c)}}%
   \putbox{3.68 in}{5.46in}{1.68}{$\lambda_{\text{\textit{on}}}$ = $10^8$~Hz  }%
   \putbox{4.5 in}{2.14in}{1.68}{No shift in $f_c$}%
   \putbox{4.41 in}{1.80in}{1.68}{for $T^{-1}_{\text{\textit{on}}}~<~ \lambda_{\text{\textit{on}}}$}%
   \putbox{6.71 in}{1.78in}{1.68}{for $T^{-1}_{\text{\textit{on}}}~>~ \lambda_{\text{\textit{on}}}$}%
   \putbox{6.87 in}{2.12in}{1.68}{$f_c~ \approx~ T^{-1}_{\text{\textit{on}}}$}%
   } 
   } 
  \vspace{-0.05cm} 
\caption { The switched bias RTS noise PSD for single transistor, calculated from derived model given in Eq. (\ref{eq:Son}) [MATLAB simulation], with variable duty cycle ($D$) to demonstrate (a) Equal shift in $f_c$ for equal ON time ($T_{\text{\textit{on}}}$) = 5$\times10^{-11}~s$, (b) Different shift in $f_c$ with variable $T_{\text{\textit{on}}}$ and equal time period ($T$) or switching frequency ($f_s$ = $1/T$), and (c) Relationship between $f_c$ and $T_{\text{\textit{on}}}$.  }

\label{cornerfchange}                                                                  
\end{figure*}


\par 
Initial PTO of ON state of the device would be zero only if enough OFF time is available for emission, which can be achieved by reducing the duty cycle of the biasing signal as per Eq.~(\ref{eq:pon}). Reduction in duty cycle results in the higher noise reduction as per Eq.~(\ref{eq:Son}). 
In Fig. \ref{fig:RTS} comparison is shown between stationary PSD and switched PSD based on the models proposed in \cite{paper:TianTHESIS, paper:gockenNOISE}, and this work. The model, shown in \cite{paper:TianTHESIS} predicts 33 dB reduction in the RTS noise PSD as compared to the standard model, whereas in \cite{paper:gockenNOISE} a  noise reduction of 86 dB is claimed. As predicted in Eq. (\ref{eq:Sn}), if $n$ increases from 2 to 4 (the duty cycle changes from 50\% to 25\%) additional 9 dB reduction is obtained which further increases up to 27 dB for $n$ =16. 
\par Based on the above analysis, a circuit level noise reduction technique is proposed for the source follower transistor in standard CMOS image sensors. The switching activity makes the  output discrete in nature, which is not suitable for image sensor applications. Complementary switches are used to obtain a continuous output in~\cite{paper:KohCOMLEMNTSWITCH}. The two MOSFET switches are periodically switched between strong inversion and accumulation or depletion regions which lead to a reduction in the low-frequency noise. The model in [9] is limited to two switches, which is extended to multiple transistors configuration (in this work), with higher noise reduction.

\section{Detailed circuit implementation}
\label{sec:CirDes}
\par 
A source follower (SF) or buffer stage is implemented as a test circuit, using multiple transistors with a variable duty cycle switched biasing. The source follower is used to show the reduction of $1/f$ noise reduction in an active pixel sensor (APS). 
A conventional APS employs three or more transistors where all transistors, except the SF, act like a switch. The switches mostly contribute to the thermal noise, while the SF contributes to the low-frequency noise. The SF low-frequency noise is becoming more evident as transistor sizes are becoming smaller with technology scaling. The thermal noise in the pixel can be reduced using CDS \cite{paper:_Enz_CDS_chop}. In CDS, first a reset signal is sampled and after the integration period of the photons, the signal is sampled. In CDS, the reset sample is subtracted from the signal and since the thermal noise components of both signals are correlated, the overall thermal noise gets reduced. However due to the increased time period of sampling the reset and the signal, the low-frequency noise becomes uncorrelated and thus the overall low-frequency noise increases. Thus, the $1/f$ noise remains as a dominant source of the noise at the imager output and must be reduced to improve the overall signal-to-noise ratio.  

\par 
The implemented circuit is shown in Fig. \ref{fig:SF}(b) in which multiple transistors are used, to implement single stage buffer/amplifier, to reduce the low-frequency noise. The purpose of multiple and identical stages is to retrieve the continuous signal at the output of the system which uses switched biasing. The output signal is discrete in nature due to switched biasing and the time for which signal is available at the output is decreased with a decrease in the duty cycle. To retain the continuous nature of the output, multiple stages are used, as shown in Fig. \ref{fig:SF}(b), which is an example of switched bias multiple transistor source follower. Individual source followers, among the multiple paths, are selected periodically for a small interval of time 
as shown in Fig. \ref{fig:SF}(c). Thus, among multiple paths, only one path from input to the output is ON at a given time. Since one path is always ON between the input and the output, the output remains continuous. If each transistor is ON for $T_{\text{\textit{on}}}$ { = $T/n$} time in a time period $T$, to achieve a continuous output, `$n$' number of stages would be required such that one stage is ON at a time. Thus, in this paper `$n$' is number of stages and used for $T_{\text{\textit{on}}} = T/n$, duty cycle ($D$) = 100 $\times$ $T/n$ $\%$.    

\begin{figure*} [ht]
\centering

\def\putbox#1#2#3#4{\makebox[0in][l]{\makebox[#1][l]{}\raisebox{\baselineskip}[0in][0in]{\raisebox{#2}[0in][0in]{\scalebox{#3}{#4}}}}}
\def\rightbox#1{\makebox[0in][r]{#1}}
\def\centbox#1{\makebox[0in]{#1}}
\def\topbox#1{\raisebox{-0.60\baselineskip}[0in][0in]{#1}}
\def\midbox#1{\raisebox{-0.20\baselineskip}[0in][0in]{#1}}
   \scalebox{0.5081}{
   \normalsize
   \parbox{13.6875in}{
   \includegraphics[scale=1.96812]{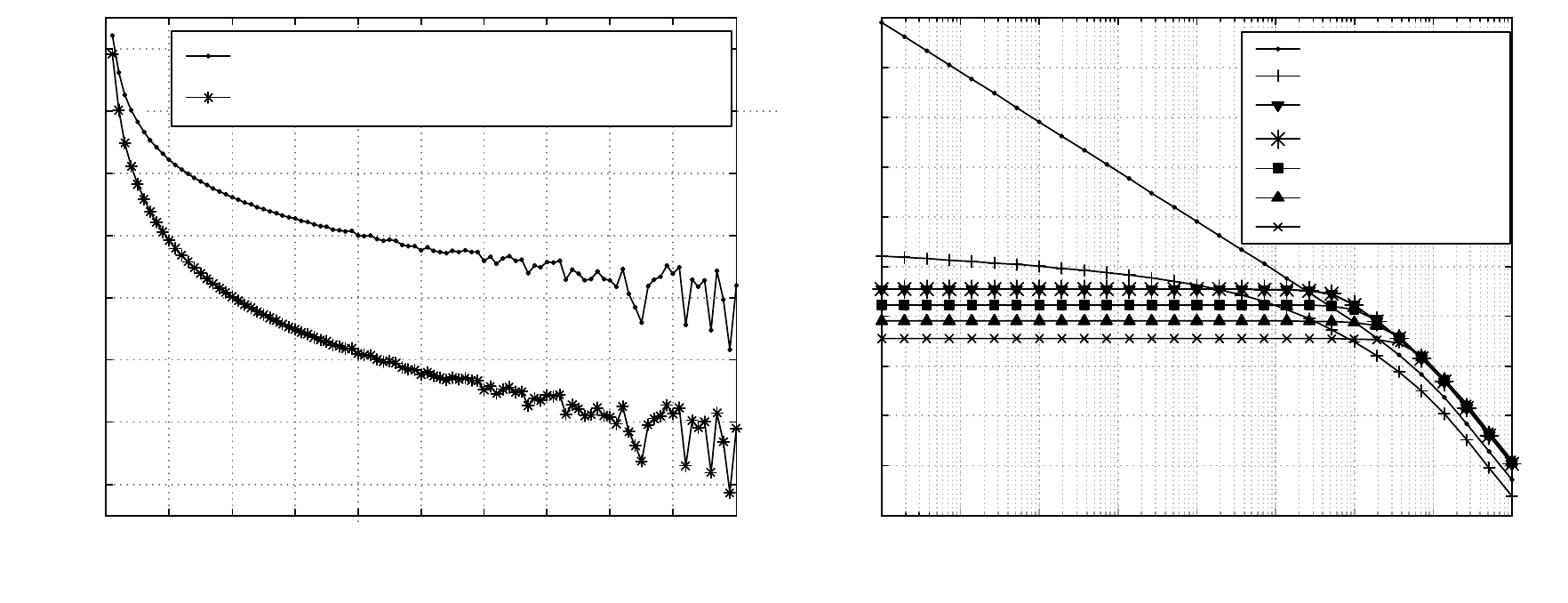}\\
   \putbox{0.87 in}{0.88in}{1.7}{\rightbox{\topbox{-160}}}%
   \putbox{0.87 in}{1.28in}{1.7}{\rightbox{\topbox{-150}}}%
   \putbox{0.87 in}{1.72in}{1.7}{\rightbox{\topbox{-140}}}%
   \putbox{0.87 in}{2.16in}{1.7}{\rightbox{\topbox{-130}}}%
   \putbox{0.87 in}{2.60in}{1.7}{\rightbox{\topbox{-120}}}%
   \putbox{0.87 in}{3.04in}{1.7}{\rightbox{\topbox{-110}}}%
   \putbox{0.87 in}{3.48in}{1.7}{\rightbox{\topbox{-100}}}%
   \putbox{0.85 in}{3.92in}{1.7}{\rightbox{\topbox{-90}}}%
   \putbox{0.85 in}{4.36in}{1.7}{\rightbox{\topbox{-80}}}%
   \putbox{0.85 in}{4.80in}{1.7}{\rightbox{\topbox{-70}}}%
   \putbox{0.85 in}{5.24in}{1.7}{\rightbox{\topbox{-60}}}%
   \putbox{0.95 in}{0.62in}{1.7}{\centbox{\topbox{$10^{6}$}}}%
   \putbox{1.66 in}{0.62in}{1.7}{\centbox{\topbox{$10^{7}$}}}%
   \putbox{2.37 in}{0.62in}{1.7}{\centbox{\topbox{$10^{8}$}}}%
   \putbox{3.08 in}{0.62in}{1.7}{\centbox{\topbox{$10^{9}$}}}%
   \putbox{3.81 in}{0.62in}{1.7}{\centbox{\topbox{$10^{10}$}}}%
   \putbox{4.5 in}{0.62in}{1.7}{\centbox{\topbox{$10^{11}$}}}%
   \putbox{5.2 in}{0.62in}{1.7}{\centbox{\topbox{$10^{12}$}}}%
   \putbox{5.89 in}{0.62in}{1.7}{\centbox{\topbox{$10^{13}$}}}%
   \putbox{6.6 in}{0.62in}{1.7}{\centbox{\topbox{$10^{14}$}}}%
   \putbox{4.7 in}{4.89in}{1.7}{\midbox{Stationary PSD}}%
   \putbox{4.7 in}{4.67in}{1.7}{\midbox{\cite{paper:TianTHESIS}}}%
   \putbox{4.7 in}{4.41in}{1.7}{\midbox{\cite{paper:gockenNOISE}}}%
   \putbox{4.7 in}{4.11in}{1.6}{\midbox{$n$ = 2, $D$ = 50\%}}%
   \putbox{4.7 in}{3.85in}{1.6}{\midbox{$n$ = 4, $D$ = 25\%}}%
   \putbox{4.7 in}{3.59in}{1.6}{\midbox{$n$ = 8, $D$ = 12.5\%}}%
   \putbox{4.7 in}{3.33in}{1.6}{\midbox{$n$ = 16, $D$ = 6.25\%}}%
   \putbox{-5.86 in}{0.60in}{1.7}{\centbox{\topbox{0}}}%
   \putbox{-5.34 in}{0.60in}{1.7}{\centbox{\topbox{20}}}%
   \putbox{-4.81 in}{0.60in}{1.7}{\centbox{\topbox{40}}}%
   \putbox{-4.23 in}{0.60in}{1.7}{\centbox{\topbox{60}}}%
   \putbox{-3.67 in}{0.60in}{1.7}{\centbox{\topbox{80}}}%
   \putbox{-3.15 in}{0.60in}{1.7}{\centbox{\topbox{100}}}%
   \putbox{-2.59 in}{0.60in}{1.7}{\centbox{\topbox{120}}}%
   \putbox{-2.03 in}{0.60in}{1.7}{\centbox{\topbox{140}}}%
   \putbox{-1.47 in}{0.60in}{1.7}{\centbox{\topbox{160}}}%
   \putbox{-0.92 in}{0.60in}{1.7}{\centbox{\topbox{180}}}%
   \putbox{-0.36 in}{0.60in}{1.7}{\centbox{\topbox{200}}}%
   \putbox{-6 in}{1.03in}{1.7}{\rightbox{\midbox{-220}}}%
   \putbox{-6 in}{1.58in}{1.7}{\rightbox{\midbox{-210}}}%
   \putbox{-6 in}{2.13in}{1.7}{\rightbox{\midbox{-200}}}%
   \putbox{-6 in}{2.68in}{1.7}{\rightbox{\midbox{-190}}}%
   \putbox{-6 in}{3.24in}{1.7}{\rightbox{\midbox{-180}}}%
   \putbox{-6 in}{3.78in}{1.7}{\rightbox{\midbox{-170}}}%
   \putbox{-6 in}{4.33in}{1.7}{\rightbox{\midbox{-160}}}%
   \putbox{-6 in}{4.89in}{1.7}{\rightbox{\midbox{-150}}}%
   \putbox{-3.18 in}{0.2 in}{2.10}{\centbox{\topbox{`$n$'}}}%
   \putbox{-4.75 in}{4.45in}{1.7}{\midbox{PSD: Variable duty cycle with single stage }}%
   \putbox{-4.75 in}{4.82in}{1.7}{\midbox{PSD: Variable number of stages}}%
   \putbox{0.05 in}{3 in}{1.8}{\rotatebox{-270}{\centbox{$1/f$ Noise PSD [dB]}}}%
   \putbox{-6.85 in}{3 in}{1.8}{\rotatebox{-270}{\centbox{RTS Noise PSD [dB]}}}%
   \putbox{2.85 in}{0.15 in}{2.10}{\midbox{Frequency [Hz]}}%
   \putbox{-5.61 in}{1.08in}{2.10}{\midbox{\textbf{(a)}}}%
   \putbox{1.14 in}{1.02in}{2.10}{\midbox{\textbf{(b)}}}%
   \putbox{-2.65 in}{5.44in}{1.9}{\rightbox{\topbox{$T$ = $10^{-11}~$s ($f_s$ = 100 GHz)}}}%
   } 
   } 
\caption{(a) RTS noise PSD of single stage with variable duty cycle ($S_{\lambda,{on}}(\omega)$) and multiple stages with variable duty cycle ($n \times S_{\lambda,{on}}(\omega)$) \cite{paper:Mypaper}, (b) Comparison of total $1/f$ noise PSD ($n\times S_{1/f}(\omega)$) at the output [MATLAB simulation], between the existing noise models and this work with variable duty cycle ($D$) and multiple (`$n$' number of) stages.}
\label{fig:nSw} 
\end{figure*}

\par 
For switched biasing, a programmable ring counter is used to generate non-overlapping clocks with a desired duty cycle, as shown in Fig. \ref{fig:SF}(b) and (c). The high level of the clock corresponds to the input voltage ($V_{in}$). The programmable ring counter can be configured to select a variable number of stages. The ring counter activates one transistor at a time. The frequency of the counter can be controlled by an external clock, thus, allowing for varying the duty cycle of each transistor, externally. As only one transistor is ON at a time in the input to the output path, the noise generated from each stage is non-correlated. Thus, the overall noise at the output is equal to the addition of the noise power of each stage transistor. This non-correlated noise sampling is one of the major keys behind the total noise reduction.                      
\par

\section{Results}
\label{sec:Res}
\subsection{Simulation results
		} 
\label{subsec:Sim_res}
Equation (\ref{eq:Sn}) is evaluated using MATLAB to demonstrate reduction in the $1/f$ noise.  
It is evident from the evaluated results, shown in Fig. \ref{cornerfchange}, that the RTS noise PSD experiences an equal shift in $f_c$ for equal $T_{\text{\textit{on}}}$ with different switching frequencies.  The RTS noise PSD is plotted for stationary noise (DC biasing) \cite{paper:vanderzeril88}, switched noise PSD based on stationary noise model \cite{paper:TianTHESIS}, and the noise PSD from the model presented in this paper. Figure \ref{cornerfchange}(a) shows the RTS noise power evaluated from Eq. (\ref{eq:Son}) for different values of time period ($T$) and $T_{\text{\textit{on}}}$ ( = $T/n$). For calculations the value of $\lambda_{\text{\textit{on}}}$ has been chosen as $10^{8}$~Hz. The RTS noise is plotted for $T$ varying from $10^{-6}$~$s$ to $10^{-10}$~$s$ and for each value of $T$,  $T_{\text{\textit{on}}}$ is kept equal to 5$\times10^{-11}$~$s$ by adjusting the value of $n$. From Fig. \ref{cornerfchange}, it can be observed that for each case the $f_c$ is shifted to approximately $T^{-1}_{\text{\textit{on}}}$. 
\par 
In Fig. \ref{cornerfchange}(b) the RTS noise evaluated from Eq. (\ref{eq:Son})) is plotted for varying values of $T_{\text{\textit{on}}}$ from 5$\times10^{-7}$~$s$ to 5$\times10^{-10}$~$s$ while keeping $T$ = $10^{-5}$~$s$. As predicted in section \ref{sec:fnoise_red}, for the noise plots with $T^{-1}_{\text{\textit{on}}}$ $<$ $\lambda_{\text{\textit{on}}}$, the corner frequency doesn't shift and remains approximately equal to $\lambda_{\text{\textit{on}}}$ ($10^{8}$~Hz). For these plots the noise power reduces by `$n$' time as compared to stationary noise. Hence, there is no reduction in the RTS noise. While for the plots with $T^{-1}_{\text{\textit{on}}}$ $>$ $\lambda_{\text{\textit{on}}}$ the corner frequency shifts to $T^{-1}_{\text{\textit{on}}}$ and the reduction in noise is higher than `$n$' time as compared to stationary noise. A higher shift in the $f_c$ accompanies a higher reduction in the low-frequency noise. It can be concluded from the results that the noise reduction depends on the $T_{\text{\textit{on}}}$ rather than $T$ or $f_s$. Thus, the same reduction can be achieved by switching the transistor at a comparatively lower frequency by decreasing the duty cycle of switching signal. The relation between $T^{-1}_{\text{\textit{on}}}$ and $f_c$ is plotted in Fig. \ref{cornerfchange}(c). It can be seen that $f_c$ stays at $\lambda_{\text{\textit{on}}}$ = ($10^{8}$~Hz), for all values of $T^{-1}_{\text{\textit{on}}}$ less than $\lambda_{\text{\textit{on}}}$, while shifts to approx. $T^{-1}_{\text{\textit{on}}}$ for all values of $T^{-1}_{\text{\textit{on}}}$ greater than $\lambda_{\text{\textit{on}}}$.\par
Equation (\ref{eq:Son}), thus shows an increased RTS noise reduction with decrease in duty cycle (100$T/n$) or increase in value of '$n$'.
The reduction in the noise is more than 10log($n$). Hence, when multiple transistors are used with the same duty cycle to ensure a continuous output, the total noise (with non-correlated noise components from multiple stages) would be less than what is obtained by using a single transistor which is ON all the time. This was shown using a mathematical model in our previous paper \cite{paper:Mypaper}. The RTS noise PSD of single stage ($S_{\lambda,{on}}(\omega)$) and multiple stages ($n \times S_{\lambda,{on}}(\omega)$) with variable duty cycle is shown in  Fig.~\ref{fig:nSw}(a). The figure shows that the RTS noise PSD decreases with decrease in duty cycle or increasing the number of stages.

\par
\begin{figure*}
\centering
\def\putbox#1#2#3#4{\makebox[0in][l]{\makebox[#1][l]{}\raisebox{\baselineskip}[0in][0in]{\raisebox{#2}[0in][0in]{\scalebox{#3}{#4}}}}}
\def\rightbox#1{\makebox[0in][r]{#1}}
\def\centbox#1{\makebox[0in]{#1}}
\def\topbox#1{\raisebox{-0.60\baselineskip}[0in][0in]{#1}}
\def\midbox#1{\raisebox{-0.20\baselineskip}[0in][0in]{#1}}
   \scalebox{0.53112}{
   \normalsize
   \parbox{16.1 in}{
   \includegraphics[scale=1.88281]{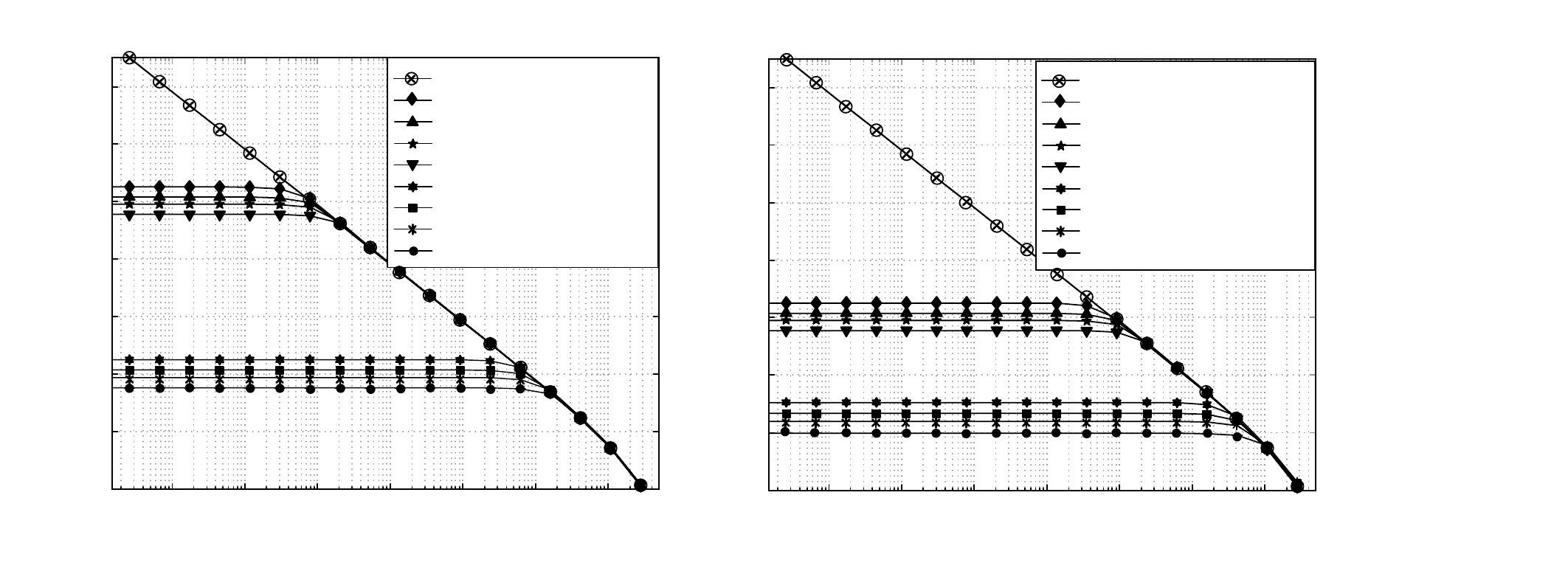}\\
   \putbox{0 in}{1 in}{1.44}{(\textbf{b})}%
   \putbox{-0.53 in}{0.80in}{1.44}{-90}%
   \putbox{-0.53 in}{1.39in}{1.44}{-80}%
   \putbox{-0.53 in}{1.97in}{1.44}{-70}%
   \putbox{-0.53 in}{2.56in}{1.44}{-60}%
   \putbox{-0.53 in}{3.14in}{1.44}{-50}%
   \putbox{-0.53 in}{3.73in}{1.44}{-40}%
   \putbox{-0.53 in}{4.34in}{1.44}{-30}%
   \putbox{-0.53 in}{4.90in}{1.44}{-20}%
   \putbox{-6.34 in}{0.6in}{1.44}{$ 10^2 $}%
   \putbox{-5.6 in}{0.6in}{1.44}{$ 10^3 $}%
   \putbox{-4.87 in}{0.6in}{1.44}{$ 10^4 $}%
   \putbox{-4.12 in}{0.6in}{1.44}{$ 10^5 $}%
   \putbox{-3.38 in}{0.6in}{1.44}{$ 10^6 $}%
   \putbox{-2.68 in}{0.6in}{1.44}{$ 10^7 $}%
   \putbox{-1.89 in}{0.6in}{1.44}{$ 10^8 $}%
   \putbox{-6.8 in}{1 in}{1.44}{(\textbf{a})}%
   \putbox{-7.27 in}{0.82in}{1.44}{-90}%
   \putbox{-7.27 in}{1.41in}{1.44}{-80}%
   \putbox{-7.27 in}{1.99in}{1.44}{-70}%
   \putbox{-7.27 in}{2.58in}{1.44}{-60}%
   \putbox{-7.27 in}{3.17in}{1.44}{-50}%
   \putbox{-7.27 in}{3.76in}{1.44}{-40}%
   \putbox{-7.27 in}{4.34in}{1.44}{-30}%
   \putbox{-7.27 in}{4.93in}{1.44}{-20}%
   \putbox{-3.52 in}{4.75in}{1.20}{$T$ = $10^{-3}$~s, $n$ = 2, $D$ = 50$\%$}%
   \putbox{0.35 in}{0.6in}{1.44}{$ 10^2 $}%
   \putbox{1.09 in}{0.6in}{1.44}{$ 10^3 $}%
   \putbox{1.83 in}{0.6in}{1.44}{$ 10^4 $}%
   \putbox{2.58 in}{0.6in}{1.44}{$ 10^5 $}%
   \putbox{3.32in}{0.6in}{1.44}{$ 10^6 $}%
   \putbox{4.06in}{0.6in}{1.44}{$ 10^7 $}%
   \putbox{4.80in}{0.6in}{1.44}{$ 10^8 $}%
   \putbox{-3.55 in}{4.52in}{1.20}{$T$ = $10^{-3}$~s, $n$ = 4, $D$ = 25$\%$}%
   \putbox{-3.55 in}{4.31in}{1.20}{$T$ = $10^{-3}$~s, $n$ = 5, $D$ = 20$\%$}%
   \putbox{-3.55 in}{4.08in}{1.20}{$T$ = $10^{-3}$~s, $n$ = 6, $D$ = 16.6$\%$}%
   \putbox{-3.55 in}{3.86in}{1.20}{$T$ = $10^{-6}$~s, $n$ = 2, $D$ = 50$\%$}%
   \putbox{-3.55 in}{3.63in}{1.20}{$T$ = $10^{-6}$~s, $n$ = 4, $D$ = 25$\%$}%
   \putbox{-3.55 in}{3.43in}{1.20}{$T$ = $10^{-6}$~s, $n$ = 5, $D$ = 20$\%$}%
   \putbox{-3.55 in}{3.20in}{1.20}{$T$ = $10^{-6}$~s, $n$ = 6, $D$ = 16.6$\%$}%
   \putbox{-3.55 in}{4.99in}{1.20}{Stationary noise}%
   \putbox{3.0 in}{4.75in}{1.20}{$T$ = $10^{-5}$~s, $n$ = 2, $D$ = 50$\%$}%
   \putbox{3.0 in}{4.52in}{1.20}{$T$ = $10^{-5}$~s, $n$ = 4, $D$ = 25$\%$}%
   \putbox{3.0 in}{4.32in}{1.20}{$T$ = $10^{-5}$~s, $n$ = 5, $D$ = 20$\%$}%
   \putbox{3.0 in}{4.09in}{1.20}{$T$ = $10^{-5}$~s, $n$ = 6, $D$ = 16.6$\%$}%
   \putbox{3.0 in}{3.87in}{1.20}{$T$ = 2$\times10^{-7}$~s, $n$ = 2, $D$ = 50$\%$}%
   \putbox{3.0 in}{3.64in}{1.20}{$T$ = 2$\times10^{-7}$~s, $n$ = 4, $D$ = 25$\%$}%
   \putbox{3.0 in}{3.43in}{1.20}{$T$ = 2$\times10^{-7}$~s, $n$ = 5, $D$ = 20$\%$}%
   \putbox{3.0 in}{3.20in}{1.20}{$T$ = 2$\times10^{-7}$~s, $n$ = 6, $D$ = 16.6$\%$}%
   \putbox{3.0 in}{4.97in}{1.20}{Stationary noise}%
   \putbox{-7.65 in}{1.87in}{1.71}{\rotatebox{-270}{$1/f$ Noise PSD $V^2/Hz $[dB]}}%
   \putbox{-0.95 in}{1.74in}{1.71}{\rotatebox{-270}{$1/f$ Noise PSD $V^2/Hz $[dB]}}%
   \putbox{-5.1 in}{0.21in}{1.71}{\rotatebox{-360}{Frequency [Hz]}}%
   \putbox{1.7 in}{0.15in}{1.71}{Frequency [Hz]}%
   } 
   } 
   \vspace{- 0.1 in} 
   \caption{(a) The $1/f$ noise PSD [$n\times S_{1/f}(\omega)$] at the output, calculated from derived model (Eq. (\ref{eq:Sformula})) [MATLAB simulation], for switched biasing with a variable duty cycle ($D$) and multiple (`$n$' number of) stages with (a) $T$ = $10^{-3}~s$ and $T$ = $10^{-6}~s$, (b) $T$ = $10^{-5}~s$ and $T$ = 2$\times10^{-7}~s$.}
\label{fig:FNOISE_new}                                                                  
\vspace{- 0.5cm}
\end{figure*}

 
In Fig. \ref{fig:nSw}(b), the $1/f$ noise PSD evaluated from Eq. (\ref{eq:Sformula}) with the multiple stages is compared with the noise PSD from the standard model (stationary noise model), and other models reported in \cite{paper:TianTHESIS} and \cite{paper:gockenNOISE}. Greater noise reduction is obtained, as shown in the figure, as compared to other models while the output is still continuous in nature. As the number of stages is increased from 2 to 4, an additional $1/f$ noise reduction of 3 dB is observed, while the RTS noise reduction obtained is 9 dB, as in Fig. \ref{fig:RTS}. When the number of stages is increased from 2 to 16 an additional $1/f$ noise reduction of 10 dB is achieved. The model, given in \cite{paper:TianTHESIS} claimed to obtain the lower noise power at higher frequencies, due to incorrect boundary conditions used. The $1/f$ noise evaluated from Eq. (\ref{eq:Sn}) is shown in Fig. \ref{fig:FNOISE_new} for varying time period from 1 KHz to 5 MHz and the number of transistor stages from 2 to 6. The stationary noise has also been plotted in the figure for comparison. At 100~Hz sampling frequency and $f_s$ = 1 KHz, the $1/f$ noise reduction obtained is 17.44 dB, 19.2 dB, 20.45 dB, 21.42 dB, and 22.21 dB for 2, 3, 4, 5, and 6 stages, respectively. This increases to 54.74 dB, 56.6 dB, 58.05 dB, 59.12 dB, and 60.21 dB for 2, 3, 4, 5, and 6 stages, respectively for 100~Hz sampling frequency and $f_s$ = 5 MHz. As the $T_{\text{\textit{on}}}$ decreases with increase in the number of stages and $f_s$ both, it can be concluded from the results, evaluated from the derived equations, that $1/f$ noise reduction depends on the $T_{\text{\textit{on}}}$ and independent of $f_s$ when $T_{\text{\textit{on}}}$ is kept constant.

\begin{figure*}[ht!]
\def\putbox#1#2#3#4{\makebox[0in][l]{\makebox[#1][l]{}\raisebox{\baselineskip}[0in][0in]{\raisebox{#2}[0in][0in]{\scalebox{#3}{#4}}}}}
\def\rightbox#1{\makebox[0in][r]{#1}}
\def\centbox#1{\makebox[0in]{#1}}
\def\topbox#1{\raisebox{-0.60\baselineskip}[0in][0in]{#1}}
\def\midbox#1{\raisebox{-0.20\baselineskip}[0in][0in]{#1}}
   \scalebox{0.3261}{
   \normalsize
   \parbox{22.3281in}{
   \includegraphics[scale=3.06655]{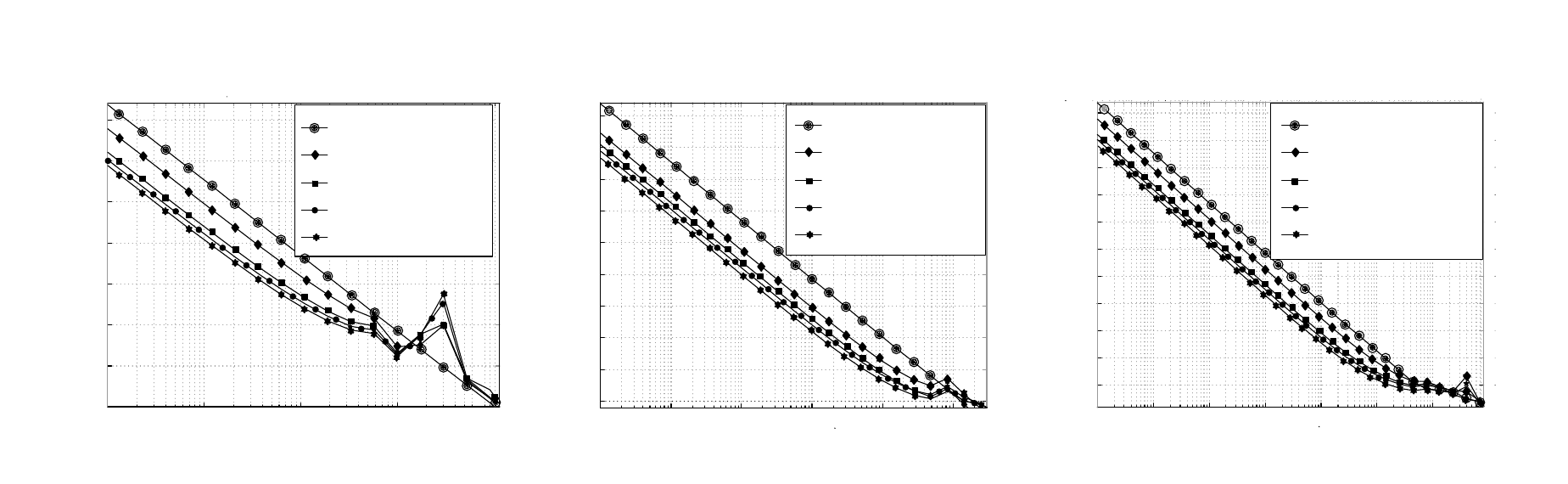}\\
   \putbox{-10.3 in}{1.02in}{2.16}{-135}%
   \putbox{-10.3 in}{1.61in}{2.16}{-130}%
   \putbox{-10.3 in}{2.20in}{2.16}{-125}%
   \putbox{-10.3 in}{2.79in}{2.16}{-120}%
   \putbox{-10.3 in}{3.39in}{2.16}{-115}%
   \putbox{-10.3 in}{3.98in}{2.16}{-110}%
   \putbox{-10.3 in}{4.57in}{2.16}{-105}%
   \putbox{-7.96 in}{0.25in}{2.16}{Frequency [Hz]}%
   \putbox{-9.57 in}{5.67in}{2.16}{$T$ = $10^{-3}~s$ ($f_s$ = 1~KHz)}%
   \putbox{-3.17 in}{1.14in}{2.16}{-145}%
   \putbox{-3.17 in}{1.59in}{2.16}{-140}%
   \putbox{-3.17 in}{2.05in}{2.16}{-135}%
   \putbox{-3.17 in}{2.51in}{2.16}{-130}%
   \putbox{-3.17 in}{2.97in}{2.16}{-125}%
   \putbox{-3.17 in}{3.43in}{2.16}{-120}%
   \putbox{-3.17 in}{3.89in}{2.16}{-115}%
   \putbox{-3.17 in}{4.34in}{2.16}{-110}%
   \putbox{-3.17 in}{4.80in}{2.16}{-105}%
   \putbox{-3.17 in}{5.26in}{2.16}{-100}%
   \putbox{4.01 in}{1.37in}{2.16}{-150}%
   \putbox{4.01 in}{1.76in}{2.16}{-145}%
   \putbox{4.01 in}{2.16in}{2.16}{-140}%
   \putbox{4.01 in}{2.55in}{2.16}{-135}%
   \putbox{4.01 in}{2.94in}{2.16}{-130}%
   \putbox{4.01 in}{3.33in}{2.16}{-125}%
   \putbox{4.01 in}{3.72in}{2.16}{-120}%
   \putbox{4.01 in}{4.12in}{2.16}{-115}%
   \putbox{4.01 in}{4.51in}{2.16}{-110}%
   \putbox{4.01 in}{4.91in}{2.16}{-105}%
   \putbox{4.01 in}{5.30in}{2.16}{-100}%
   \putbox{-0.46 in}{0.25in}{2.16}{Frequency [Hz]}%
   \putbox{6.79 in}{0.22in}{2.16}{Frequency [Hz]}%
   \putbox{0.37 in}{0.71in}{2.16}{$10^3$}%
   \putbox{1.38 in}{0.70in}{2.16}{$10^4$}%
   \putbox{2.42 in}{0.70in}{2.16}{$10^5$}%
   \putbox{-0.63 in}{0.71in}{2.16}{$10^2$}%
   \putbox{-5.69 in}{0.71in}{2.16}{$10^3$}%
   \putbox{-4.28 in}{0.70in}{2.16}{$10^4$}%
   \putbox{6.93 in}{0.71in}{2.16}{$10^3$}%
   \putbox{7.76 in}{0.71in}{2.16}{$10^4$}%
   \putbox{8.6 in}{0.71in}{2.16}{$10^5$}%
   \putbox{9.4 in}{0.72in}{2.16}{$10^6$}%
   \putbox{6.14 in}{0.70in}{2.16}{$10^2$}%
   \putbox{-7.03 in}{0.71in}{2.16}{$10^2$}%
   \putbox{-8.48 in}{0.71in}{2.16}{$10^1$}%
   \putbox{-1.72 in}{0.70in}{2.16}{$10^1$}%
   \putbox{5.34 in}{0.69in}{2.16}{$10^1$}%
   \putbox{-6.37 in}{5.05in}{1.80}{Stationary Noise PSD}%
   \putbox{-6.37 in}{4.66in}{1.80}{$n$ = 2, $D$ = 50 $\%$}%
   \putbox{-10.73 in}{1.50in}{2.16}{\rotatebox{-270}{$1/f$ Noise PSD $V^2/Hz $[dB]}}%
   \putbox{-9.66 in}{0.71in}{2.16}{$10^0$}%
   \putbox{-2.58 in}{0.71in}{2.16}{$10^0$}%
   \putbox{4.59 in}{0.71in}{2.16}{$10^0$}%
   \putbox{-10.3 in}{5.16in}{2.16}{-100}%
   \putbox{-2.26 in}{1.25in}{2.16}{\textbf{(b)}}%
   \putbox{-9.38 in}{1.29in}{2.16}{\textbf{(a)}}%
   \putbox{4.91 in}{1.29in}{2.16}{\textbf{(c)}}%
   \putbox{-6.37 in}{3.85in}{1.80}{$n$ = 5, $D$ = 20 $\%$}%
   \putbox{-6.37 in}{4.25in}{1.80}{$n$ = 4, $D$ = 25 $\%$}%
   \putbox{-6.37 in}{3.47in}{1.80}{$n$ = 6, $D$ = 16.6 $\%$}%
   \putbox{-2.48 in}{5.67in}{2.16}{$T$ = $10^{-5}~s$ ($f_s$ = 100~KHz)}%
   \putbox{4.73 in}{5.67in}{2.16}{$T$ = $10^{-6}~s$ ($f_s$ = 1~MHz)}%
   \putbox{-3.69 in}{1.50in}{2.16}{\rotatebox{-270}{$1/f$ Noise PSD $V^2/Hz $[dB]}}%
   \putbox{3.39 in}{1.52in}{2.16}{\rotatebox{-270}{$1/f$ Noise PSD $V^2/Hz $[dB]}}%
   \putbox{0.77 in}{5.09in}{1.80}{Stationary Noise PSD}%
   \putbox{0.77 in}{4.70in}{1.80}{$n$ = 2, $D$ = 50 $\%$}%
   \putbox{0.77 in}{3.89in}{1.80}{$n$ = 5, $D$ = 20 $\%$}%
   \putbox{0.77 in}{4.28in}{1.80}{$n$ = 4, $D$ = 25 $\%$}%
   \putbox{0.77 in}{3.51in}{1.80}{$n$ = 6, $D$ = 16.6 $\%$}%
   \putbox{7.8 in}{5.09in}{1.80}{Stationary Noise PSD}%
   \putbox{7.8 in}{4.70in}{1.80}{$n$ = 2, $D$ = 50 $\%$}%
   \putbox{7.8 in}{3.89in}{1.80}{$n$ = 5, $D$ = 20 $\%$}%
   \putbox{7.8 in}{4.28in}{1.80}{$n$ = 4, $D$ = 25 $\%$}%
   \putbox{7.8 in}{3.51in}{1.80}{$n$ = 6, $D$ = 16.6 $\%$}%
   } 
   } 
   \vspace{- 0.1 in} 
  \caption{Simulation results (from Spectre, Cadence) for the $1/f$ noise PSD with multiple (`$n$' number of) stages and duty cycle ($D$) with (a) $T$ = $10^{-3}~s$, (a) $T$ = $10^{-5}~s$, and (c) $T$ = $10^{-6}~s$. \vspace{- 0.1 in}}
  \label{CADMSRNT}                                                         \end{figure*}

The noise PSD for multiple stage configuration, generated by the periodic steady state (PSS) and pnoise analysis from Spectre simulator by Cadence IC-615, is shown in Fig. \ref{CADMSRNT}. The MOSFET model used is Star-Hspice level 49 (BSIM3V3.2) with UMC 0.18um Mixed-mode/RFCMOS 1.8V 1P6M P-sub twin-well CMOS salicide process. The different graphs (a), (b), and (c) show the noise results for switching frequencies of 1 KHz, 100 KHz, and 1 MHz respectively. According to simulation results, a total reduction in the noise power is $10log(n)$ dB for $n$ stage configuration. It can be seen in the graphs that corner frequency shifts to $f_s$ and the noise reduction is almost constant, which is different from the results obtained from Eq. (\ref{eq:Sformula}). 

\begin{figure*}
\centering
\def\putbox#1#2#3#4{\makebox[0in][l]{\makebox[#1][l]{}\raisebox{\baselineskip}[0in][0in]{\raisebox{#2}[0in][0in]{\scalebox{#3}{#4}}}}}
\def\rightbox#1{\makebox[0in][r]{#1}}
\def\centbox#1{\makebox[0in]{#1}}
\def\topbox#1{\raisebox{-0.60\baselineskip}[0in][0in]{#1}}
\def\midbox#1{\raisebox{-0.20\baselineskip}[0in][0in]{#1}}
   \scalebox{0.738459}{
   \normalsize
   \parbox{9.47917in}{
   \includegraphics[scale=1.35417]{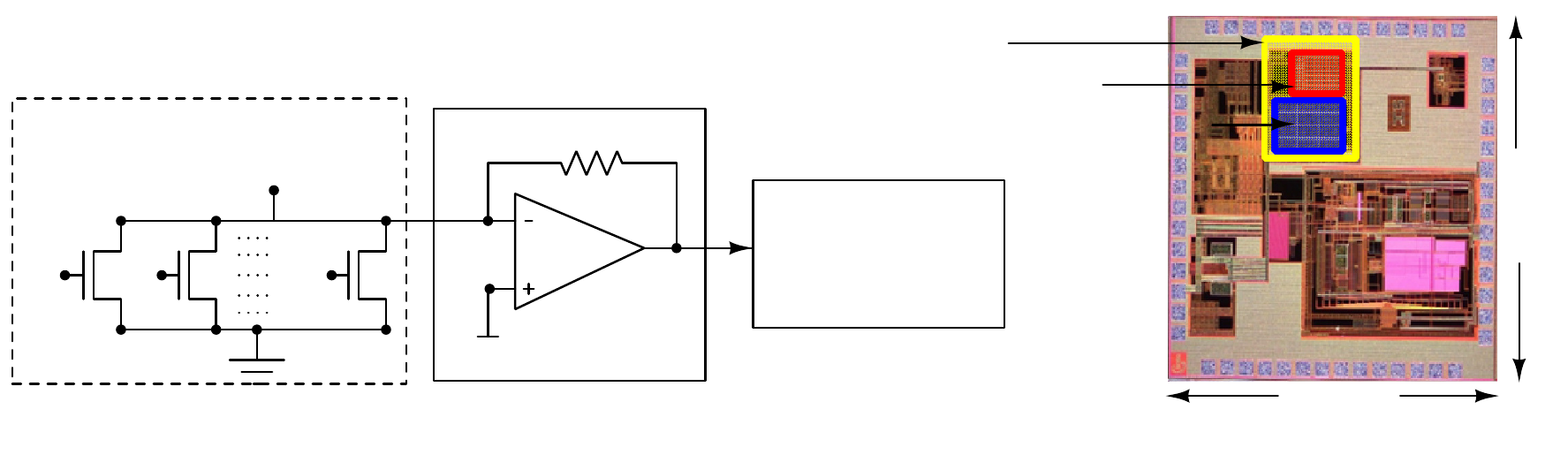}\\
   \putbox{-1.97 in}{0.67in}{1.20}{\midbox{$V_{Ref}$}}%
   \putbox{0.11 in}{1.26in}{1.20}{\midbox{Dynamic Signal}}%
   \putbox{0.38 in}{1.04in}{1.20}{\midbox{Analyzer}}%
   \putbox{0.44 in}{1.49in}{1.20}{\midbox{SR785}}%
   \putbox{-1.21 in}{2.59in}{1.20}{\centbox{\midbox{SR 570 Low Noise}}}%
   \putbox{-1.12 in}{2.37in}{1.20}{\centbox{\midbox{Current preamplifier}}}%
   \putbox{-4.66 in}{1.20in}{1.20}{\midbox{$clk_1$}}%
   \putbox{-4.1 in}{1.20in}{1.20}{\midbox{$clk_2$}}%
   \putbox{-1.35 in}{2.07in}{1.20}{\midbox{$R_{gain}$}}%
   \putbox{-2.54 in}{0.18in}{1.20}{\midbox{(a)}}%
   \putbox{0.93 in}{2.37in}{1.20}{\midbox{Ring counter}}%
   \putbox{5.67 in}{2.13in}{1.20}{\midbox{Multi-Transistor stages}}%
   \putbox{3.33 in}{0.17in}{1.20}{\midbox{(b)}}%
   \putbox{3.15 in}{0.45in}{1.20}{\midbox{1.5 mm}}%
   \putbox{4.57 in}{1.92in}{1.20}{\rotatebox{-90}{\midbox{1.5 mm}}}%
   \putbox{-3.09 in}{1.21in}{1.20}{\midbox{$clk_n$}}%
   \putbox{-3.25 in}{1.87in}{1.20}{\midbox{$V_{bias}$}}%
   \putbox{0.93 in}{2.6in}{1.2}{\midbox{DUT}}%
   \putbox{-3.64 in}{2.43in}{1.56}{\midbox{DUT}}%
   } 
   } 
   \vspace{-\baselineskip} 
\caption{[Color online] (a) Measurement setup for design under test (DUT) (b) Chip microphotograph. }
\label{setup}                                                                  
\end{figure*}
\begin{figure*}[ht!]
     \centering

\def\putbox#1#2#3#4{\makebox[0in][l]{\makebox[#1][l]{}\raisebox{\baselineskip}[0in][0in]{\raisebox{#2}[0in][0in]{\scalebox{#3}{#4}}}}}
\def\rightbox#1{\makebox[0in][r]{#1}}
\def\centbox#1{\makebox[0in]{#1}}
\def\topbox#1{\raisebox{-0.60\baselineskip}[0in][0in]{#1}}
\def\midbox#1{\raisebox{-0.20\baselineskip}[0in][0in]{#1}}
   \scalebox{0.518519}{
   \normalsize
   \parbox{13.5in}{
   \includegraphics[scale=1.92857]{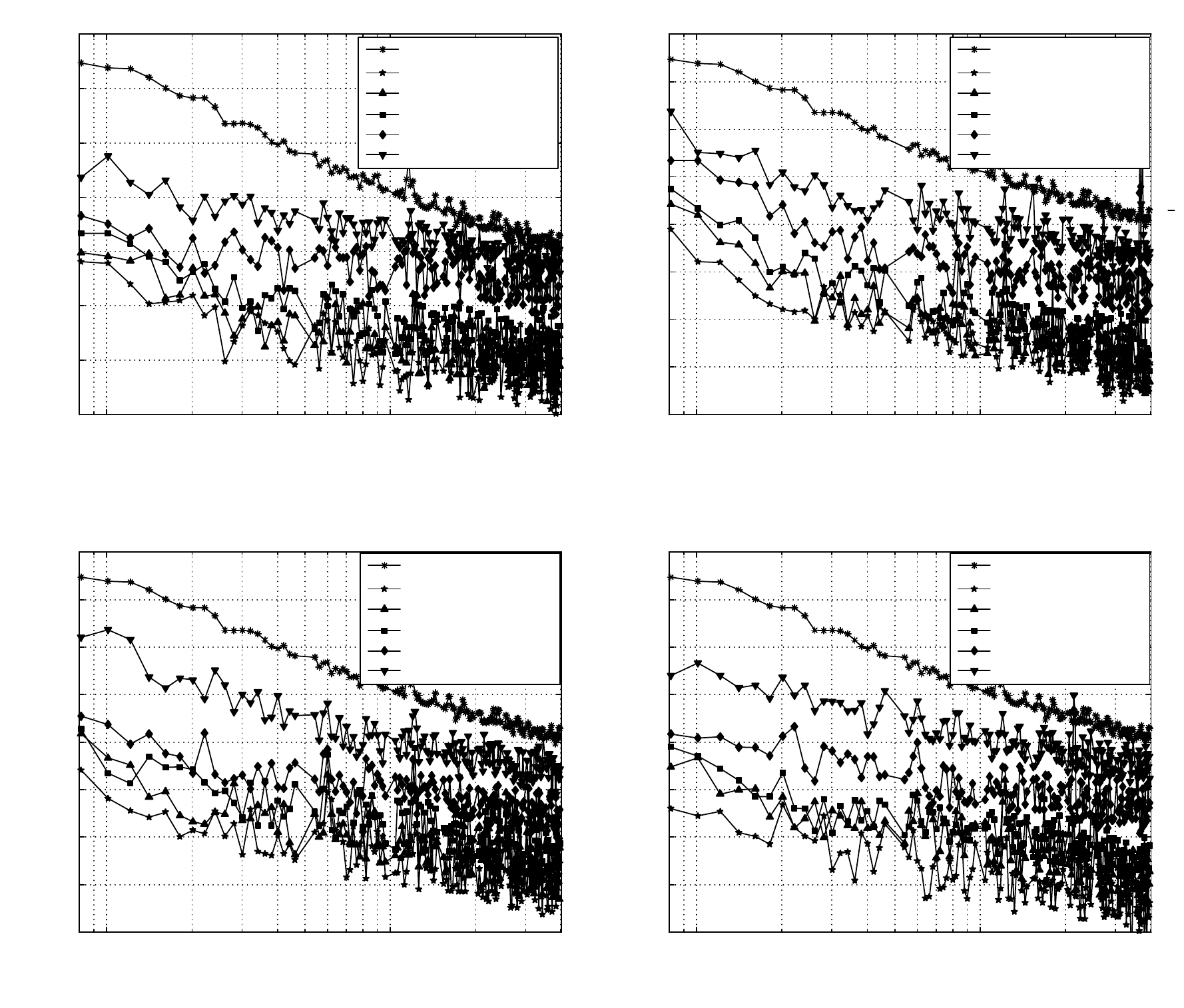}\\
   \putbox{-6.3 in}{6.84in}{1.44}{-110}%
   \putbox{-6.3 in}{7.47in}{1.44}{-105}%
   \putbox{-6.3 in}{8.10in}{1.44}{-100}%
   \putbox{-6.24 in}{8.72in}{1.44}{-95}%
   \putbox{-6.24 in}{9.35in}{1.44}{-90}%
   \putbox{-6.24 in}{9.98in}{1.44}{-85}%
   \putbox{-6.24 in}{10.61in}{1.44}{-80}%
   \putbox{-6.24 in}{11.24in}{1.44}{-75}%
   \putbox{0.53 in}{6.84in}{1.44}{-115}%
   \putbox{0.53 in}{7.39in}{1.44}{-110}%
   \putbox{0.53 in}{7.93in}{1.44}{-105}%
   \putbox{0.53 in}{8.49in}{1.44}{-100}%
   \putbox{0.6 in}{9.04in}{1.44}{-95}%
   \putbox{0.6 in}{9.59in}{1.44}{-90}%
   \putbox{0.6 in}{10.14in}{1.44}{-85}%
   \putbox{0.6 in}{10.69in}{1.44}{-80}%
   \putbox{0.6 in}{11.24in}{1.44}{-75}%
   \putbox{-6.3 in}{0.84in}{1.44}{-115}%
   \putbox{-6.3 in}{1.39in}{1.44}{-110}%
   \putbox{-6.3 in}{1.93in}{1.44}{-105}%
   \putbox{-6.3 in}{2.49in}{1.44}{-100}%
   \putbox{-6.24 in}{3.04in}{1.44}{-95}%
   \putbox{-6.24 in}{3.59in}{1.44}{-90}%
   \putbox{-6.24 in}{4.14in}{1.44}{-85}%
   \putbox{-6.24 in}{4.69in}{1.44}{-80}%
   \putbox{-6.24 in}{5.24in}{1.44}{-75}%
   \putbox{0.53 in}{0.84in}{1.44}{-115}%
   \putbox{0.53 in}{1.39in}{1.44}{-110}%
   \putbox{0.53 in}{1.93in}{1.44}{-105}%
   \putbox{0.53 in}{2.49in}{1.44}{-100}%
   \putbox{0.53 in}{3.04in}{1.44}{-95}%
   \putbox{0.53 in}{3.59in}{1.44}{-90}%
   \putbox{0.53 in}{4.14in}{1.44}{-85}%
   \putbox{0.53 in}{4.69in}{1.44}{-80}%
   \putbox{0.53 in}{5.24in}{1.44}{-75}%
   \putbox{-5.59 in}{0.59in}{1.44}{\midbox{10}}%
   \putbox{-2.4 in}{0.59in}{1.44}{\midbox{100}}%
   \putbox{-4.44 in}{0.26in}{1.80}{\midbox{Frequency [Hz]}}%
   \putbox{1.24 in}{0.59in}{1.44}{\midbox{10}}%
   \putbox{4.44 in}{0.59in}{1.44}{\midbox{100}}%
   \putbox{2.39 in}{0.26in}{1.80}{\midbox{Frequency [Hz]}}%
   \putbox{1.24 in}{6.59in}{1.44}{\midbox{10}}%
   \putbox{4.44 in}{6.59in}{1.44}{\midbox{100}}%
   \putbox{2.39 in}{6.26in}{1.80}{\midbox{Frequency [Hz]}}%
   \putbox{-5.67 in}{6.59in}{1.44}{\midbox{10}}%
   \putbox{-2.48 in}{6.59in}{1.44}{\midbox{100}}%
   \putbox{-4.52 in}{6.26in}{1.80}{\midbox{Frequency [Hz]}}%
   \putbox{-6.5 in}{7.85in}{1.80}{\rotatebox{-270}{\midbox{ $1/f$ Noise PSD $V^2/Hz $[dB]}}}%
   \putbox{-6.54 in}{1.78in}{1.80}{\rotatebox{-270}{\midbox{ $1/f$ Noise PSD $V^2/Hz $[dB]}}}%
   \putbox{0.41 in}{7.85in}{1.80}{\rotatebox{-270}{\midbox{ $1/f$ Noise PSD $V^2/Hz $[dB]}}}%
   \putbox{0.37 in}{1.78in}{1.80}{\rotatebox{-270}{\midbox{ $1/f$ Noise PSD $V^2/Hz $[dB]}}}%
   \putbox{-5.85 in}{11.45in}{1.80}{\midbox{$T$ = $10^{-3}~s$ ($f_s$ = 1~KHz)}}%
   \putbox{0.9 in}{11.47in}{1.80}{\midbox{$T$ = $10^{-5}~s$ ($f_s$ = 100~KHz)}}%
   \putbox{0.98 in}{5.47in}{1.80}{\midbox{$T$ = 2$\times10^{-7}~s$ ($f_s$ = 5~MHz)}}%
   \putbox{-5.93 in}{5.45in}{1.80}{\midbox{$T$ = $10^{-6}~s$ ($f_s$ = 1~MHz)}}%
   \putbox{-5.58 in}{7.16in}{1.68}{\textbf{(a)}}%
   \putbox{1.32 in}{7.10in}{1.68}{\textbf{(b)}}%
   \putbox{5.54 in}{1.12in}{1.68}{\textbf{(c)}}%
   \putbox{1.21 in}{1.16in}{1.68}{\textbf{(d)}}%
   \putbox{-2.08 in}{11.09in}{1.44}{\midbox{Stationary PSD}}%
   \putbox{-2.08 in}{10.85in}{1.44}{\midbox{$n$ = 6, $D$ = 16.6~\%}}%
   \putbox{-2.08 in}{10.60in}{1.44}{\midbox{$n$ = 5, $D$ = 20~\%}}%
   \putbox{-2.08 in}{10.37in}{1.44}{\midbox{$n$ = 4, $D$ = 25~\%}}%
   \putbox{-2.08 in}{10.12in}{1.44}{\midbox{$n$ = 3, $D$ = 33.3~\%}}%
   \putbox{-2.08 in}{9.88in}{1.44}{\midbox{$n$ = 2, $D$ = 50~\%}}%
   \putbox{4.77 in}{11.09in}{1.44}{\midbox{Stationary PSD}}%
   \putbox{4.77 in}{10.85in}{1.44}{\midbox{$n$ = 6, $D$ = 16.6 \%}}%
   \putbox{4.77 in}{10.60in}{1.44}{\midbox{$n$ = 5, $D$ = 20 \%}}%
   \putbox{4.77 in}{10.37in}{1.44}{\midbox{$n$ = 4, $D$ = 25 \%}}%
   \putbox{4.77 in}{10.12in}{1.44}{\midbox{$n$ = 3, $D$ = 33.3 \%}}%
   \putbox{4.77 in}{9.88in}{1.44}{\midbox{$n$ = 2, $D$ = 50 \%}}%
   \putbox{-2.08 in}{5.11in}{1.44}{\midbox{Stationary PSD}}%
   \putbox{-2.08 in}{4.87in}{1.44}{\midbox{$n$ = 6, $D$ = 16.6 \%}}%
   \putbox{-2.08 in}{4.62in}{1.44}{\midbox{$n$ = 5, $D$ = 20 \%}}%
   \putbox{-2.08 in}{4.39in}{1.44}{\midbox{$n$ = 4, $D$ = 25 \%}}%
   \putbox{-2.08 in}{4.14in}{1.44}{\midbox{$n$ = 3, $D$ = 33.3 \%}}%
   \putbox{-2.08 in}{3.90in}{1.44}{\midbox{$n$ = 2, $D$ = 50 \%}}%
   \putbox{4.77 in}{5.11in}{1.44}{\midbox{Stationary PSD}}%
   \putbox{4.77 in}{4.87in}{1.44}{\midbox{$n$ = 6, $D$ = 16.6 \%}}%
   \putbox{4.77 in}{4.62in}{1.44}{\midbox{$n$ = 5, $D$ = 20 \%}}%
   \putbox{4.77 in}{4.39in}{1.44}{\midbox{$n$ = 4, $D$ = 25 \%}}%
   \putbox{4.77 in}{4.14in}{1.44}{\midbox{$n$ = 3, $D$ = 33.3 \%}}%
   \putbox{4.77 in}{3.90in}{1.44}{\midbox{$n$ = 2, $D$ = 50 \%}}%
   } 
   } 
  \vspace{- 0.2 in} 
   \caption{Measurement results for the switched bias $1/f$ Noise PSD for single stage configuration with a variable duty cycle ($D$) or continuous ON time ($T_{\text{\textit{on}}}$) with (a) $T$ = $10^{-3}~s$ , (a) $T$ = $10^{-5}~s$, (a) $T$ = $10^{-6}~s$, and (d) $T$ = 5$\times10^{-7}~s$. }
   \label{onestage}                                                                  
    \end{figure*}   

\subsection{Measurement results}
\label{subsec:mea_res} 

The measurement setup is shown in Fig. \ref{setup}(a) while the microphotograph of the test circuit, fabricated in 180~nm standard 6M2P CMOS process, is shown in Fig. \ref{setup}(b). The DUT has employed an array of nMOS transistors and a configurable ring counter to generate periodic switched biasing signals for turning the MOS transistors ON and OFF, periodically. The devices used in test circuit are nMOS transistors with $W=1~\mu m$  and $L = 1~\mu m$. The target application of the proposed method is to reduce the $1/f$ noise in CMOS image sensors. The device dimensions for noise characterization are chosen to be small to maximize the fill factor of a pixel and maintain the smaller area of the pixel for higher spatial resolution.    
\par 
The measurement setup is similar to that reported in \cite{paper:blaum01, paper:wei01}. The low noise SR570 current preamplifier is used to provide the biasing current to the test circuit and to amplify the noise power. As shown in Fig. \ref{setup}(b) the drain to source voltage $V_{DS}$ (or $V_{bias}$) of test transistors, is equal to V$_{ref}$ of the preamplifier. The value of $V_{DS}$ can thus be set to a desired value by varying V$_{ref}$ to keep transistors in saturation region during ON state. The noise current generated from the test circuit is amplified by $R_{gain}$, to produce the noise voltage at the output of the preamplifier. The current to voltage sensitivity of the preamplifier was set at 10 $\mu A/V$ to get amplified noise voltage at the output. The input referred noise of current preamplifier is as low as 5 fA/$\sqrt{Hz}$, which makes it quite suitable for precise noise measurement. The preamplifier noise signal voltage is fed into the SR785 dynamic signal analyzer (DSA). DSA plots the  Fourier transform of the noise voltage signal coming from the preamplifier. The input referred noise of DSA is as low as -160 dBV$_{rms}$/$\sqrt{Hz}$ and the maximum bandwidth is 102.4 KHz. The noise PSD is plotted with 400 line FFT resolution and 400 Hz span to set 1~Hz frequency resolution. The span of the dynamic signal analyzer was kept as 400~Hz with 400 line FFT. As the maximum frequency was 400~Hz, the sampling rate is 1024 samples/sec with 1024 number of points in a time record. Dynamic signal analyzer is employed with a digital anti-aliasing filter with variable cut off frequency. The cut off frequency of the filter depends on the span of FFT spectrum. The time record length also depends on the span of the spectrum. After sampling, an anti-aliasing filter is used which suitably rejects the  out-of-band spectral components. In the measurement cutoff frequency of the digital low-pass filter contained is 512 Hz with flat response upto 400~Hz and roll-off from 400 Hz to 512 Hz. To improve the measurement accuracy, each noise PSD curve is plotted after taking an RMS average of 100 measured samples.

\begin{table*}[ht!]

\caption{Summary of Anticipated Reduction in The Noise power for Multiple  Stage Transistor from The Measured Noise of Switched Biased single Transistor}

\label{tab:Table 1}
  \begin{tabular}{p{0.8cm}||p{1.1cm}||p{0.8cm}|p{0.8cm}|p{0.8cm}|p{0.8cm}||p{0.8cm}|p{0.8cm}|p{0.8cm}|p{0.8cm}||p{0.8cm}|p{0.8cm}|p{0.8cm}|p{0.8cm}||p{0.8cm}|p{0.8cm}|p{0.8cm}|p{0.8cm}}
   
  \hline
    
  $^\ast$$\textbf{$\textbf{$\textbf{\textit{T}}$}$=}$ & {\textbf{\textit{$\infty$~(dc)}}} & \multicolumn{4}{c||}{\textbf{$\textbf{10}^{\textbf{-3}}~\textbf{\textit{s}}$}}  & \multicolumn{4}{c||}{\textbf{$\textbf{10}^{\textbf{-5}}~\textbf{\textit{s}}$}}  & \multicolumn{4}{c||}{\textbf{$\textbf{10}^{\textbf{\textbf{-6}}}~\textbf{\textit{s}}$}}  & \multicolumn{4}{c}{\textbf{$\textbf{5}\times\textbf{10}^{\textbf{-7}}~\textbf{\textit{s}}$}}  \\  \hline
	
 {$^\star${$\textbf{\textit{n}}$} =}  &  \textbf{ 1}  &  \textbf{6} &  \textbf{4}   &  \textbf{3} &   \textbf{2}  &  \textbf{6} &  \textbf{4}   &  \textbf{3} &   \textbf{2}   &  \textbf{6} &  \textbf{4}   &  \textbf{3} &   \textbf{2}   &  \textbf{6} &  \textbf{4}   &  \textbf{3} &   \textbf{2}   \\ \hline  \hline
  
  {$^\bullet$$\textbf{{\textit{D}}}$} =   &  \textbf{ 100}  &  \textbf{16.6} &  \textbf{ 25}   &  \textbf{33.3} &   \textbf{50}  & \textbf{16.6}  &  \textbf{25}   &  \textbf{33.3}  &   \textbf{50}  & \textbf{16.6}  &  \textbf{25}   &  \textbf{33.3} & \textbf{50}    & \textbf{16.6}   & \textbf{25}     &  \textbf{33.3} &  \textbf{50}   \\ \hline  \hline
   
   {$^\diamond$$\textbf{\textit{f}}_{\textbf{\textit{m}}}$}$\downarrow$  &  \multicolumn{17}{c} 
 {\begin{tabular}[c]{@{}c@{}}  {\textbf{$\textbf{1/\textit{f}}$ Noise power of multistage configuration in \textbf{\textit{V}$^\textbf{\textit{2}}$/\textit{Hz}} [dB]}} \\  \textbf{= Switched bias \textbf{1/\textit{f}} noise of single stage + 10log(\textbf{100/\textit{D}})} \end{tabular}}     \\
   \hline
   10 & -78.0 & -90.3 & -87.3  & -87.7 & -82.2 & -91.4 & -88.2 & -83.5  & -84.4 & -93.1 & -92.2 & -88.4 & -83.1  & -94.9  & -90.4  & -89.8 & -83.6 \\ \hline
   40 & -85.2 & -94.6 & -92.3  & -89.9 & -90.0 & -96.8 & -95.3 & -94.0  & -91.5 & -97.0 & -78.0 & -95   & -90.0  & -97.9  & -95.1  & -94.7 & -91.1 \\ \hline
   130 & -88.1 & -96.5 & -94.2  & -92.2 & -90.9 & -101  & -96.2 & -93.4  & -91.5 & -98.8 & -78.0 & -94.8 & -91.1  & -103   & -96.8  & -96.5 & -91.5 \\ \hline
   270 & -92.9 & -96.6 & -96.0  & -94.5 & -94.3 & -103  & -99.2 & -95.3  & -93.4 & -100  & -78.0 & -96.3 & -94.5  & -105   & -100   & -97.8 & -93   \\ \hline
   380 & -93.6 & -98.4 & -97.1  & -95.2 & -94.0 & -104  & -102 & -97.3  & -93.7 & -102 & -78.0 & -98.2 & -97.3  & -104   & -100   & -95.2 & -94.9 \\ \hline \hline
   \end{tabular}
   
   $^\ast$$T$ = Time period; $^\star$$n$ = Number of stages;$\hspace{0.1cm}$  $^\bullet$$D$ = Duty cycle [\%]; $\hspace{0.1cm}$  $^\diamond$$f_m$ = Sampling frequency [Hz]. $\hspace{0.1cm}$       
   \end{table*}


\begin{figure*}[ht!]
\centering

\def\putbox#1#2#3#4{\makebox[0in][l]{\makebox[#1][l]{}\raisebox{\baselineskip}[0in][0in]{\raisebox{#2}[0in][0in]{\scalebox{#3}{#4}}}}}
\def\rightbox#1{\makebox[0in][r]{#1}}
\def\centbox#1{\makebox[0in]{#1}}
\def\topbox#1{\raisebox{-0.60\baselineskip}[0in][0in]{#1}}
\def\midbox#1{\raisebox{-0.20\baselineskip}[0in][0in]{#1}}
   \scalebox{0.5365}{
   \normalsize
   \parbox{13.0469in}{
   \includegraphics[scale=1.86393]{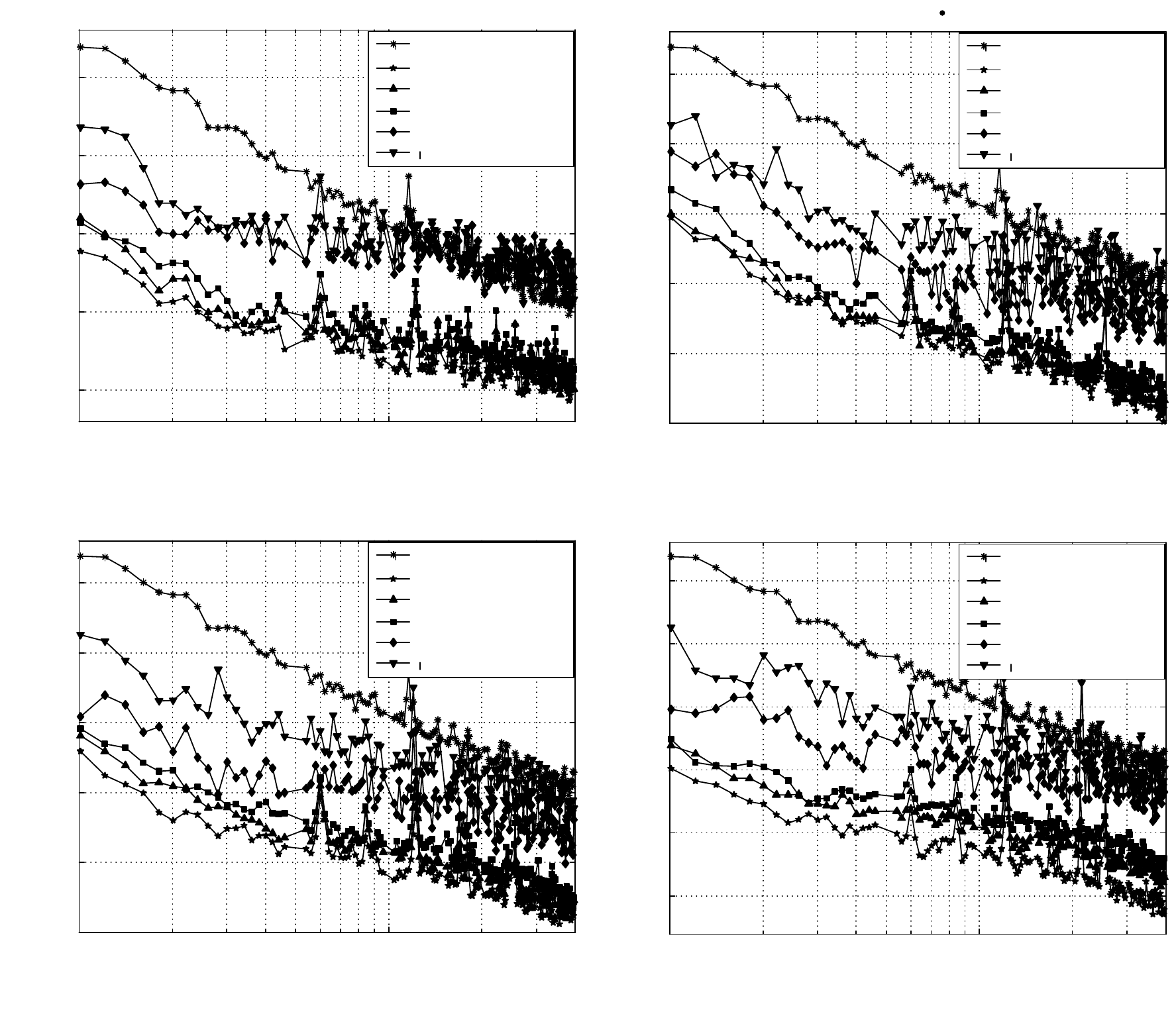}\\
   \putbox{-6.1 in}{6.92in}{1.44}{-100}%
   \putbox{-6.05 in}{7.80in}{1.44}{-95}%
   \putbox{-6.05 in}{8.68in}{1.44}{-90}%
   \putbox{-6.05 in}{9.56in}{1.44}{-85}%
   \putbox{-6.05 in}{10.44in}{1.44}{-80}%
   \putbox{0.52 in}{6.55in}{1.44}{-105}%
   \putbox{0.52 in}{7.33in}{1.44}{-100}%
   \putbox{0.59 in}{8.12in}{1.44}{-95}%
   \putbox{0.59 in}{8.91in}{1.44}{-90}%
   \putbox{0.59 in}{9.69in}{1.44}{-85}%
   \putbox{0.59 in}{10.48in}{1.44}{-80}%
   \putbox{-6.1 in}{0.82in}{1.44}{-105}%
   \putbox{-6.1 in}{1.61in}{1.44}{-100}%
   \putbox{-6.05 in}{2.39in}{1.44}{-95}%
   \putbox{-6.05 in}{3.18in}{1.44}{-90}%
   \putbox{-6.05 in}{3.96in}{1.44}{-85}%
   \putbox{-6.05 in}{4.75in}{1.44}{-80}%
   \putbox{0.52 in}{1.22in}{1.44}{-105}%
   \putbox{0.52 in}{1.93in}{1.44}{-100}%
   \putbox{0.59 in}{2.65in}{1.44}{-95}%
   \putbox{0.59 in}{3.36in}{1.44}{-90}%
   \putbox{0.59 in}{4.06in}{1.44}{-85}%
   \putbox{0.59 in}{4.78in}{1.44}{-80}%
   \putbox{-5.77 in}{11.17in}{1.80}{\midbox{$T$ = $10^{-3}~s$ ($f_s$ = 1~KHz)}}%
   \putbox{-5.77 in}{5.46in}{1.80}{\midbox{$T$ = $10^{-6}~s$ ($f_s$ = 1~MHz)}}%
   \putbox{0.96 in}{5.49in}{1.80}{\midbox{$T$ = 2$\times10^{-7}~s$ ($f_s$ = 5~MHz)}}%
   \putbox{0.96 in}{11.17in}{1.80}{\midbox{$T$ = $10^{-5}~s$ ($f_s$ = 100~KHz)}}%
   \putbox{-3.57 in}{6. in}{1.80}{\midbox{Frequency [Hz]}}%
   \putbox{2.83 in}{6.in}{1.80}{\midbox{Frequency [Hz]}}%
   \putbox{2.83 in}{0.2 in}{1.80}{\midbox{Frequency [Hz]}}%
   \putbox{-3.57 in}{0.2 in}{1.80}{\midbox{Frequency [Hz]}}%
   \putbox{-6.32 in}{7.53in}{1.80}{\rotatebox{-270}{\midbox{ $1/f$ Noise PSD $V^2/Hz $[dB]}}}%
   \putbox{-6.36 in}{1.63in}{1.80}{\rotatebox{-270}{\midbox{ $1/f$ Noise PSD $V^2/Hz $[dB]}}}%
   \putbox{0.33 in}{7.51in}{1.80}{\rotatebox{-270}{\midbox{ $1/f$ Noise PSD $V^2/Hz $[dB]}}}%
   \putbox{0.33 in}{1.61in}{1.80}{\rotatebox{-270}{\midbox{ $1/f$ Noise PSD $V^2/Hz $[dB]}}}%
   \putbox{4.75 in}{10.83in}{1.44}{\midbox{Stationary PSD}}%
   \putbox{4.75 in}{10.59in}{1.44}{\midbox{$n$ = 6, $D$ = 16.6~\%}}%
   \putbox{4.75 in}{10.34in}{1.44}{\midbox{$n$ = 5, $D$ = 20~\%}}%
   \putbox{4.75 in}{10.11in}{1.44}{\midbox{$n$ = 4, $D$ = 25~\%}}%
   \putbox{4.75 in}{9.86in}{1.44}{\midbox{$n$ = 3, $D$ = 33.3~\%}}%
   \putbox{4.75 in}{9.63in}{1.44}{\midbox{$n$ = 2, $D$ = 50~\%}}%
   \putbox{4.75 in}{5.08in}{1.44}{\midbox{Stationary PSD}}%
   \putbox{4.75 in}{4.84in}{1.44}{\midbox{$n$ = 6, $D$ = 16.6~\%}}%
   \putbox{4.75 in}{4.59in}{1.44}{\midbox{$n$ = 5, $D$ = 20 \%}}%
   \putbox{4.75 in}{4.36in}{1.44}{\midbox{$n$ = 4, $D$ = 25 \%}}%
   \putbox{4.75 in}{4.11in}{1.44}{\midbox{$n$ = 3, $D$ = 33.3 \%}}%
   \putbox{4.75 in}{3.88in}{1.44}{\midbox{$n$ = 2, $D$ = 50 \%}}%
   \putbox{-1.9 in}{5.10in}{1.44}{\midbox{Stationary PSD}}%
   \putbox{-1.9 in}{4.87in}{1.44}{\midbox{$n$ = 6, $D$ = 16.6~\%}}%
   \putbox{-1.9 in}{4.62in}{1.44}{\midbox{$n$ = 5, $D$ = 20 \%}}%
   \putbox{-1.9 in}{4.38in}{1.44}{\midbox{$n$ = 4, $D$ = 25 \%}}%
   \putbox{-1.9 in}{4.13in}{1.44}{\midbox{$n$ = 3, $D$ = 33.3 \%}}%
   \putbox{-1.9 in}{3.90in}{1.44}{\midbox{$n$ = 2, $D$ = 50 \%}}%
   \putbox{-1.9 in}{10.85in}{1.44}{\midbox{Stationary PSD}}%
   \putbox{-1.9 in}{10.62in}{1.44}{\midbox{$n$ = 6, $D$ = 16.6~\%}}%
   \putbox{-1.9 in}{10.37in}{1.44}{\midbox{$n$ = 5, $D$ = 20 \%}}%
   \putbox{-1.9 in}{10.13in}{1.44}{\midbox{$n$ = 4, $D$ = 25 \%}}%
   \putbox{-1.9 in}{9.88in}{1.44}{\midbox{$n$ = 3, $D$ = 33.3 \%}}%
   \putbox{-1.9 in}{9.65in}{1.44}{\midbox{$n$ = 2, $D$ = 50 \%}}%
   \putbox{-5.78 in}{6.44in}{1.44}{\midbox{10}}%
   \putbox{-2.4 in}{6.44in}{1.44}{\midbox{100}}%
   \putbox{0.86 in}{6.42in}{1.44}{\midbox{10}}%
   \putbox{4.24 in}{6.42in}{1.44}{\midbox{100}}%
   \putbox{0.84 in}{0.69in}{1.44}{\midbox{10}}%
   \putbox{4.22 in}{0.69in}{1.44}{\midbox{100}}%
   \putbox{5.8 in}{0.67in}{1.44}{\midbox{10}}%
   \putbox{-2.42 in}{0.67in}{1.44}{\midbox{100}}%
   } 
   } 
 
  \vspace{- 0.05 in} 
  \caption {Measurement results for the switched bias $1/f$ noise PSD for multiple (`$n$' number of) stages and a variable duty cycle ($D$) or continuous ON time ($T_{\text{\textit{on}}}$) with (a) $T$ = $10^{-3}~s$ , (a) $T$ = $10^{-5}~s$, (a) $T$ = $10^{-6}~s$, and (d) $T$ = 5$\times10^{-7}~s$. }
  \label{MultiStage}                                                                  
     \end{figure*}   
   
 \par 

The DUT has a configurable ring counter which generates non-overlapping clocks ($clk_1$, $clk_2$..., and $clk_n$) to select one MOS transistor at a time, as described in section \ref{sec:CirDes}. Non-overlapping clocks make sure that the noise components from different stages are non-correlated. The noise measurements for DC biasing are carried out with $V_{DS} = V_{GS} = 1~V$ for nMOS transistors. For switched biasing, $V_{DS} = V_{GS}$ is set to $1~V$ during ON time to keep transistors in the saturation region while $V_{GS} = 0~V$ for OFF time to keep the transistor in cutoff region. First, the noise measurements are carried out on a single transistor with constant biasing. Then measurement are carried out for single as well as multiple transistors, with switched biasing and a variable duty cycle ($ D $). 

The noise power at the output node for the circuit shown in Fig. 3(b) can be calculated by MOSFET noise current ($I_{DS_n}$) (the noise current flows from drain to source of the device) and the load resistance. The load resistance only contributes thermal noise, thus $1/f$ noise can be analyzed by measuring the noise current $I_{DS_n}$ at low frequencies. The measurement setup, shown in Fig. 8(a), is used for measuring the noise current. The noise current flows from load $R_{gain}$ to get amplified by current to voltage amplifier (SR 570), and then measured by Dynamic Signal Analyzer (DSA - SR785), as shown in Fig. 8(a). From the noise measurement results obtained by DSA, the reduction in the noise current of DUT are analyzed.
\par 
 
The measured noise power of a single stage transistor with constant and switched biasing is shown in Fig.~\ref{onestage}. The noise plots, shown in Fig. \ref{onestage}(a), (b), (c), and (d) demonstrates the reduction in the low-frequency noise power of single transistor, switched bias at $f_s$ = 1 KHz, 100 KHz, 1 MHz, and 5 MHz respectively. For each $f_s$, the duty cycle is taken as $16.6~\%$, $20~\%$, $25~\%$, and $50~\%$. 
The average reduction in $1/f$ noise power for single stage configuration with varying duty cycle, (for sampling frequency up to 40~Hz and $f_s$ = 1~kHz) evaluated from Eq. (\ref{eq:Sformula}), {using MATLAB simulation}, is 29.57 dB, when the duty cycle varies from 100~$\%$ (DC biasing) to 50~$\%$. After which when the duty cycle reduces to 25~$\%$, 20~$\%$ and 16.6~$\%$ the average noise reduction increases to 35.5 dB, 37.6 dB, and 39.11 dB, respectively.
While the measurement results (Fig.~\ref{onestage}) shows an average reduction obtained in $1/f$ noise, for $f_s$ = 1~KHz is 7.02 dB when the duty cycle varies from 100~\% to 50~\%. As the duty cycle further reduces to 25~\%, 20~\% and 16.6~\%, the noise reduction increases to 15.5~dB, 17.3~dB, and 18.72~dB respectively.
The sampling frequency has been chosen only up to 40~Hz, to show the low-frequency noise reduction.
\par
As discussed in section \ref{sec:fnoise_red}, the low-frequency noise can be reduced using variable duty cycle and multiple stages. The noise from each of these stages are non-correlated in nature; hence, the overall noise PSD at the output can be calculated by summing the noise power from an individual stage. In order to compare the noise from single stage transistor with DC biasing and a varying duty cycle switched biasing, a factor of 10log$_{10}(D)$ is added into the later. By this addition, the noise from multiple stage configuration is predicted and summarized in Table \ref{tab:Table 1}. 
\par 
The measured noise power sampled at frequency points at 10~Hz, 40~Hz, 270~Hz, and 380~Hz for varying $f_s$ and $D$, as given in Table \ref{tab:Table 1}. The noise reduction with multiple stages (calculated from the switched noise PSD of single transistor) should vary from 4.2 dB with 50~$\%$ duty cycle to 12.3 dB with 16.6~$\%$ duty cycle at 10~Hz sampling frequency, while the reduction should be 0.4 dB with 50~$\%$ duty cycle to 4.8 dB with 16.6~$\%$ duty cycle at 380 Hz sampling frequency, for $f_s$ = 1 KHz.
The random points are selected to avoid the effects of spikes generated due to power supply noise at 50 Hz and its harmonics.
It can be concluded from the summary given in Table \ref{tab:Table 1} that the noise power decreases with a decrease in the duty cycle at a given $f_s$. The same trend is seen when the $f_s$ is increased while the noise reduction decreases for higher sampling frequencies. The decrease in the noise reduction at higher sampling frequencies is due to the dominance of the thermal noise over low-frequency noise above $f_c$.           

\par 
The measured noise power of single stage nMOS with constant biasing and multiple nMOS transistors with switched biasing, for a varying duty cycle, is shown in Fig.~\ref{MultiStage}. The output noise power for 2 to 6 stages, switched with $f_s$ = 1 KHz, 100 KHz, 1 MHz, and 5 MHz are presented in Fig. \ref{MultiStage} (a), (b), (c), and (d) respectively. For the measurements, the ring counter is configured to select 2 to 6 transistor stages, as shown in Fig. \ref{fig:SF}(b). $f_s$ is varied from 1 KHz to 5 MHz for a variable number of stages. The duty cycle is varied from 50~\% to 16.6~\% for 2 to 6 transistor stages. From the measured results variation can be seen in the integrated noise reduction (for sampling frequency up to 40 Hz) from 5.9 dB, for 2 stages to 12.3 dB, for 6 stages, with $f_s$ = 1 KHz. 
Reduction in the noise is increased with increase in a number of stages (or decrease in the duty cycle) and $f_s$.
In both the cases, the noise reduction is increased due to the decrease in continuous ON time of the transistor, as predicted by Eq. (\ref{eq:Son}). 
 
\par 
By comparison, between the results shown in Table \ref{tab:Table 1} (the predicted noise power of `$n$' stages configuration from the measured noise for single stage with variable duty cycle by adding 10$log(n)$ dB) and Fig. \ref{MultiStage} (measured noise of multistage configuration), it can be seen that reduction in the noise power, with $f_s$ = 1 KHz, should be 4.2 dB for 2 transistor stages to 12.3 dB for 6 stages (at 10 Hz sampling frequency) and 0.4 dB for 2 stages to 4.8 dB for 6 stages (at 380 Hz sampling frequency). While the measured noise reduction obtained is 3.9 dB for 2 stage to 11.2 dB for 6 stage configuration (at 10 Hz sampled frequency) to 0.7 dB for 2 stages to 5.7 dB for 6 stages (at 380 Hz sampling frequency). 
The comparison shows that the low-frequency noise power of multistage configuration is approximately same as calculated from the noise of single transistor with a variable duty cycle switched biasing. The reason of the small difference in the noise powers is the presence of other circuits like ring counter and switched transistors. 

 \begin{figure}
 
\def\putbox#1#2#3#4{\makebox[0in][l]{\makebox[#1][l]{}\raisebox{\baselineskip}[0in][0in]{\raisebox{#2}[0in][0in]{\scalebox{#3}{#4}}}}}
\def\rightbox#1{\makebox[0in][r]{#1}}
\def\centbox#1{\makebox[0in]{#1}}
\def\topbox#1{\raisebox{-0.60\baselineskip}[0in][0in]{#1}}
\def\midbox#1{\raisebox{-0.20\baselineskip}[0in][0in]{#1}}
   \scalebox{0.5027}{
   \normalsize
   \parbox{6.1875in}{
   \includegraphics[scale=1.98926]{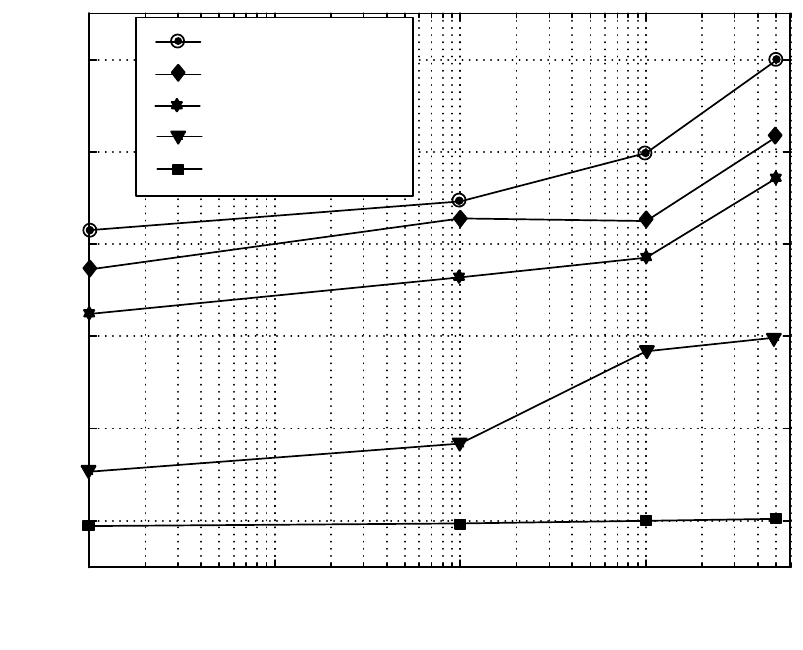}\\
   \putbox{-2.54 in}{0.39in}{1.5}{10}%
   \putbox{-1.06 in}{0.39in}{1.5}{10}%
   \putbox{0.41 in}{0.39in}{1.5}{10}%
   \putbox{1.89 in}{0.39in}{1.5}{10}%
   \putbox{-2.31 in}{0.48in}{1.2}{3}%
   \putbox{-0.83 in}{0.48in}{1.2}{4}%
   \putbox{0.62 in}{0.48in}{1.2}{5}%
   \putbox{2.12 in}{0.48in}{1.2}{6}%
   \putbox{-2.65 in}{0.98in}{1.44}{6}%
   \putbox{-2.65 in}{1.71in}{1.44}{8}%
   \putbox{-2.7 in}{2.44in}{1.5}{10}%
   \putbox{-2.7 in}{3.17in}{1.5}{12}%
   \putbox{-2.7 in}{3.91in}{1.5}{14}%
   \putbox{-2.7 in}{4.64in}{1.5}{16}%
   \putbox{-0.8 in}{0.1 in}{1.6}{Switching Frequency $f_s$ [Hz]}%
   \putbox{-3.1 in}{0.7 in}{1.6}{\rotatebox{-270}{Reduction in $1/f$ Noise PSD $V^2/Hz $[dB]}}%
   \putbox{-1.47 in}{4.77in}{1.44}{$n$ = 6, D = 16.6~\%}%
   \putbox{-1.47 in}{4.52in}{1.44}{$n$ = 5, D = 20~\%}%
   \putbox{-1.47 in}{4.26in}{1.44}{$n$ = 4, D = 25~\%}%
   \putbox{-1.47 in}{4.01in}{1.44}{$n$ = 3, D = 33.3~\%}%
   \putbox{-1.47 in}{3.77in}{1.44}{$n$ = 2, D = 50~\%}%
   } 
   } 
    \vspace{- 0.05 in} 
   \caption{Reduction in the measured $1/f$ noise power with multiple (`$n$' number of) stages, duty cycle ($D$), and switching frequency $f_s$ (or $T^{-1}$). }
   \label{Recuction}                                                                  
   \vspace{-0.5 cm} 
   \end{figure}

\par 
The average reduction in the $1/f$ noise power (for sampling frequency up to 40 Hz) is summarized in Table \ref{tab:Table 2} and Fig. \ref{Recuction}, for $f_s$ = 1 KHz, 100 KHz, 1 MHz, and 5 MHz. The $1/f$ noise power calculated by Eq. (\ref{eq:Sn}) (the noise PSD shown in Fig. \ref{fig:FNOISE_new}) predicts the average reduction varies from 24.8~dB to 31.32 dB when the number of stages vary from 2 to 6, for $f_s$ = 1 KHz.
It can be concluded from results that the $1/f$ noise power depends on the continuous ON time of the device rather than $f_s$. For higher sampling frequencies the switched bias noise PSD is approximately equal to the stationary noise PSD, as the thermal noise start dominating above the $1/f$ noise corner frequency.  
\par 
One drawback of using multiple transistors with switched biasing is that switching action of the transistors introduces ripples at the output. These ripples are generated due to clock feed-through of the overlapping capacitance present between drain and gate of the switching transistor. The ripples can be filtered out through a switch capacitor low-pass filter. The low pass filter can be placed in the column of the CMOS image sensor, which will not affect the fill factor of the pixel.
The mismatch of the source followers would increase the column FPN. The column FPN needs to be characterized with an imaging array in place. The focus of this work is to show the noise reduction obtained using multiple stages as compared to simply duty cycling a single device. In future, an imaging array with the proposed low-frequency noise reduction method will be fabricated and the effects of multiple source followers on the column FPN and other imaging performance will be characterized. \vspace{- 0.05 in}
  \begin{table}[h!]
    \centering
    \caption{Summary of The $1/f$ Noise Power (Measured) Reduction with Multistage Configuration at varying switching frequencies}
    \label{tab:Table 2}
      \begin{tabular}{c||p{0.8cm}|p{0.8cm}|p{0.8cm}|p{0.8cm}|p{0.8cm}}
      \textbf{No. of stage (\textbf{\textit{n}})} = & \textbf{6} & \textbf{5} & \textbf{4}  & \textbf{3} & \textbf{2} \\
      \hline
      $\textbf{\textit{f}}_\textbf{\textit{s}}$ $\downarrow$ & \multicolumn{5}  {c}
      {\begin{tabular}[c]{@{}c@{}}\textbf{Reduction in the Noise Power in}\\  {$\textbf{\textit{V}}^{\textbf{\textit{2}}}\textbf{\textit{/}}\textbf{\textit{Hz}}$ \textbf{[dB]}}\end{tabular}} \\
      \hline
      \hline
      1KHz & 12.3 & 11.4 & 10.4 & 7 & 5.9 \\
     \hline
     100KHz & 12.9 & 12.5 & 11.2 & 7.6 & 5.93 \\
     \hline
      1MHz & 14 & 12.5 & 11.7 & 9.7 & 6 \\
     \hline
      5MHz & 16 & 14.3 & 13.4 & 10 & 6.1
       \\
     \hline
     \hline
    \end{tabular}
    \end{table}
     \begin{figure}  [ht!]
     \begin{center}
     		\includegraphics[scale=0.15]{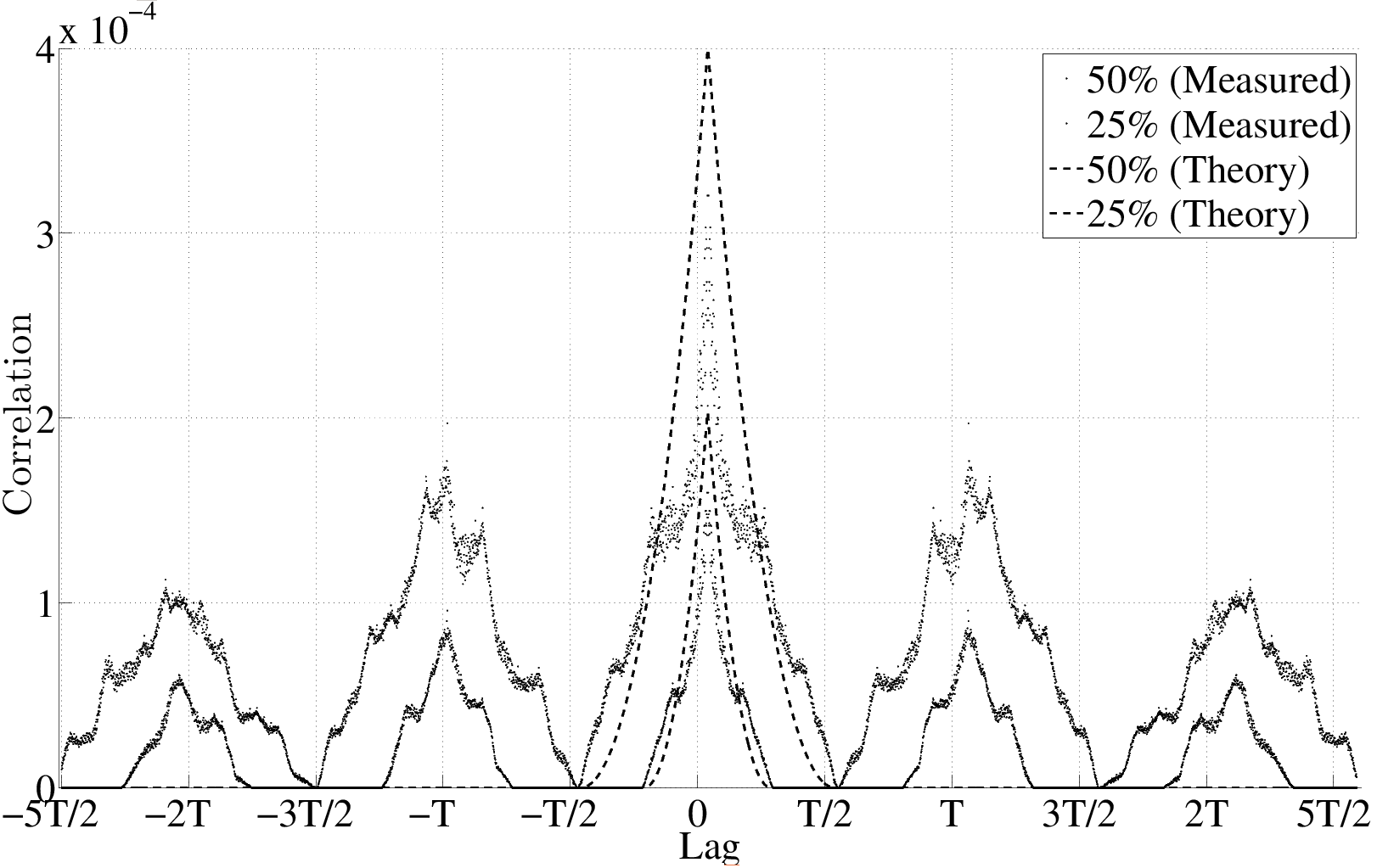}
     		\label{32K}
     \caption {ACF plots for the theoretical calculated and measurement data }
     \label{CORR} 
     \end{center}
     \end{figure} 
 
 \section{Discussion and conclusion}
\label{sec:conclusion}
In this paper, a mathematical model of the RTS noise, for a MOSFET device with time-varying biasing conditions, is presented. 
It is concluded that the RTS noise and consequently the $1/f$ noise of an MOS transistor decreases with a decrease in the $T_{\text{\textit{on}}}$. It is shown that reduction in the noise occurs due to varying correlation between trap states which is a function of $T_{\text{\textit{on}}}$ rather than switching frequency. Based on this conclusion a circuit level low-frequency noise reduction technique is presented. 
It is observed by measurement that the noise reduction which is 5.9 dB with $f_s$ = 1 KHz for 2 stage is extended up to 16 dB for 6 stages  with $f_s$ = 5 MHz.

 The nature of the low-frequency noise measurement results presented in this paper is similar to the results given in literature. There is some discrepancy in the low-frequency noise behavior between the measurement results shown in Fig. 9 and 10 and the theoretical results shown in the Fig. \ref{fig:FNOISE_new}. As per the theoretical results the noise PSD is flat for low frequencies while, the measurement results show a decreasing profile. There are two possible  reasons for the discrepancy between theoretical and measured results. Firstly, the ACF calculated in the theory and the autocorrelation obtained from the measurements are different. Secondly, there is a reappearance of $1/f$ noise in the low frequency region. 

In Fig. \ref{CORR} the theoretical ACF given by Eq. (\ref{eq:con(lt)}) is plotted for a device which is ON for $T _{on}$ time period during a time interval of $ T $. This graph has been compared with the ACF calculated for the experimentally measured noise data with 50$\%$ and 25$\%$ duty cycles. The measurement ACF is obtained from the measured noise samples. There are mainly two important differences between these ACF plots which could be the reason for the discrepancy observed between the theoretical and measured FFT plots;

     \begin{figure}
     \centering
     
     \def\putbox#1#2#3#4{\makebox[0in][l]{\makebox[#1][l]{}\raisebox{\baselineskip}[0in][0in]{\raisebox{#2}[0in][0in]{\scalebox{#3}{#4}}}}}
     \def\rightbox#1{\makebox[0in][r]{#1}}
     \def\centbox#1{\makebox[0in]{#1}}
     \def\topbox#1{\raisebox{-0.60\baselineskip}[0in][0in]{#1}}
     \def\midbox#1{\raisebox{-0.20\baselineskip}[0in][0in]{#1}}
     \begin{center}
        \scalebox{0.453324}{
        \normalsize
        \parbox{7.3385in}{
          \includegraphics[scale=0.9]{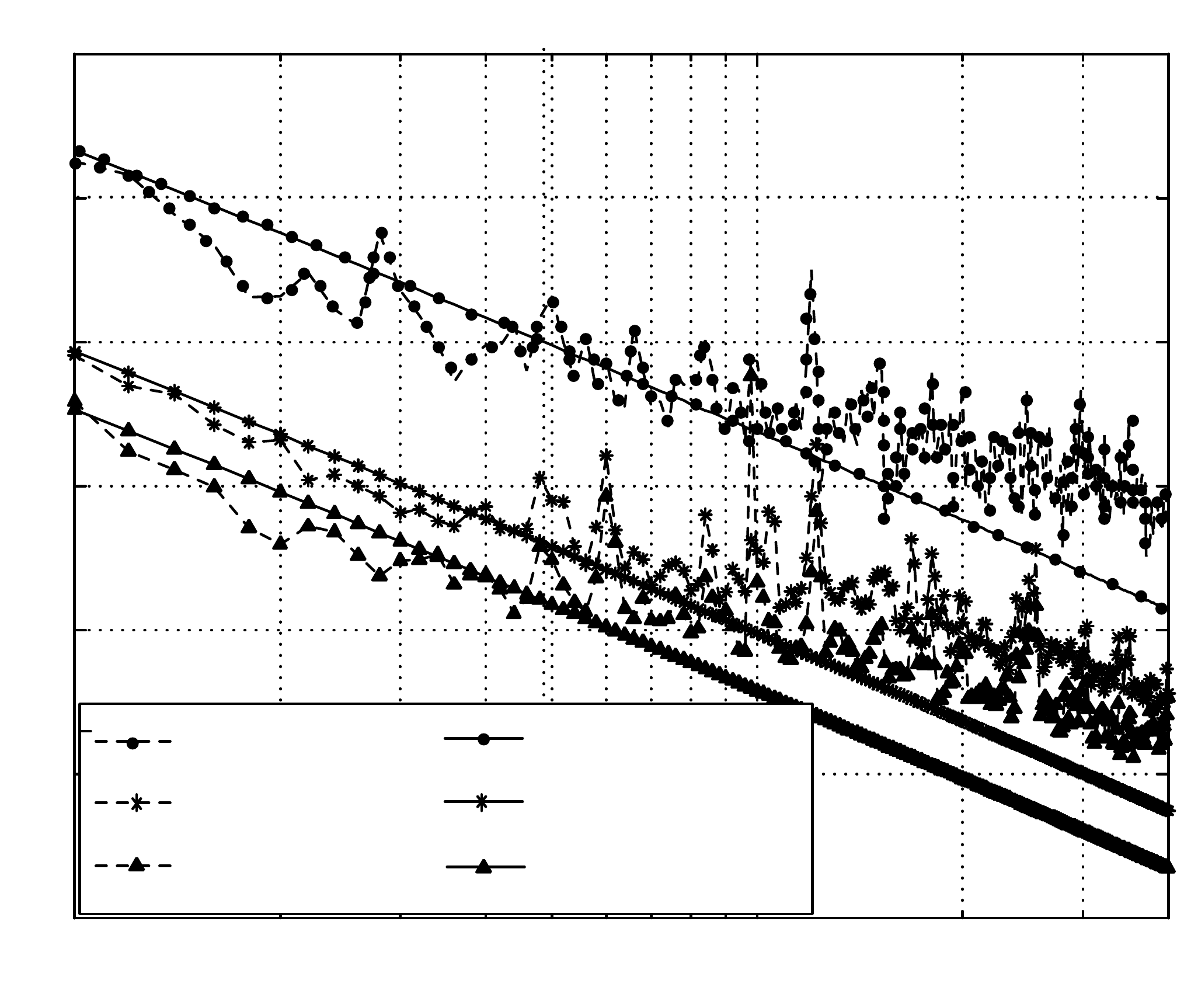}\\
          \putbox{2.1 in}{0.3in}{1.8}{10$ ^2 $}%
          \putbox{-2.1 in}{0.3 in}{1.8}{10}%
          \putbox{-3.7 in}{0.5 in}{1.8}{-110}%
          \putbox{-3.7 in}{1.35 in}{1.8}{-105}%
          \putbox{-3.7 in}{2.25 in}{1.8}{-100}%
          \putbox{-3.6 in}{3.1 in}{1.8}{-95}%
          \putbox{-3.6 in}{4 in}{1.8}{-90}%
          \putbox{-3.6 in}{4.85 in}{1.8}{-85}%
          \putbox{-3.6 in}{5.7 in}{1.8}{-80}%
          \putbox{-3.2 in}{6 in}{2.2}{$T$ = $10^{-6}~s$ ($f_s$ = 1~MHz)}%
          \putbox{-2.6 in}{1.6 in}{1.4}{Measured (2 Stage)}%
          \putbox{-2.6 in}{1.2 in}{1.4}{Measured (4 Stage)}%
          \putbox{-2.6 in}{0.8 in}{1.4}{Measured (6 Stage)}%
          \putbox{-0.4 in}{1.6 in}{1.4}{Theoritical (2 Stage)}%
          \putbox{-0.4 in}{1.2 in}{1.4}{Theoritical (4 Stage)}%
          \putbox{-0.4 in}{0.8 in}{1.4}{Theoritical (6 Stage)}%
          \putbox{-4 in}{1.5 in}{2}{\rotatebox{-270}{$ 1/f $ Noise PSD - $V^2/Hz $[dB]}}%
          \putbox{-0.7 in}{0.1 in}{2.5}{\rotatebox{-360}{Frequency [Hz]}}%
          } 
          } 
       \vspace{ -0.2   in} 
      \end{center}
       \caption {The $1/f$ noise PSD [$S'_{1/f}(\omega)$] at the output, calculated from Eq. (\ref{eq:Sformula_alter}) [MATLAB simulation], for switched biasing with a variable duty cycle ($D$) and multiple  number of stages ($n$).}
\label{fig:COMP_MSR_T}                                                             
          \end{figure}

\begin{itemize}
\item The experimentally obtained curve is not as smooth as the theoretically  obtained curve which also causes discrepancy in the nature of PSD plots.
\item It can be seen that the theoretical correlation plot has non-zero values only for time samples between $-T_ {on}$ to $T_ {on}$. However, the experimentally obtained
correlations are non-zero beyond this range.
\end{itemize}

The other reason for the the discrepancy is the reappearance of the $1/f$ noise. This effect has been discussed in \cite{paper:gockenNOISE} -  \cite{paper:reappr_conf}. Our model is based on the assumption that $\lambda_{\text{\textit{off}}}T_{\text{\textit{off}}}>>1$, as $\lambda_{\text{\textit{off}}}$ is very high for all traps. However, this condition need not be true for all traps, and thus all traps might not be affected uniformly in OFF condition during switching \cite{paper:klump_03, paper:reappr_conf}. The distribution of $\lambda_{\text{\textit{off}}}$ depends on a space dependent parameter `$m$' \cite{paper:reappr_conf}. Emission rate during off time is not uniform and can be given as:  
\begin{equation}
\lambda_\text{\textit{off}}(emission) = m.\lambda_\text{\textit{on}}(emission)
\label{eq:mon} \vspace{- 0.05 in}
\end{equation}
 The value of `$m$' needs to be $\infty$ to ensure that the trap is empty during OFF state. However `$m$' is not uniform among all traps as it depends on the location of the trap with respect to the surface of the channel~\cite{paper:klump_03, paper:reappr_conf}. The value of `$m$' is less for slow traps which are located away from the channel surface. If `$m$' for a trap is small enough to make sure that $\lambda_{\text{\textit{off}}}T_{\text{\textit{off}}}<<1$, then it's state (filled or empty) does not change with switching and hence, the noise PSD from these traps represents the noise similar to the stationary case which decreases with increase in frequency. Hence, another factor is added in the expression of the noise PSD to take account of the reappearance of the $1/f$ noise. The modified $1/f$ noise can be given as: \vspace{- 0.05 in}
\begin{equation}
S'_{1/f}(\omega) = n\int_{\lambda_{L}}^{\lambda_{H}}  S^{s}_{\lambda,on}(\omega)g(\lambda)d\lambda+\int_{\lambda'_{L}}^{\lambda'_{H}}\frac{\gamma\lambda}{{\lambda^2} + \omega^2}g(\lambda)d\lambda,
\label{eq:Sformula_alter}
\end{equation}
where $n$ is number of stages, $\lambda'_L$ and $\lambda'_H$ are the minimum and maximum transition rates, respectively among the slow traps during OFF state. $\gamma$ is the fitting factor used to match the measured results with theory. The noise PSD calculated from Eq. (\ref{eq:Sformula_alter}), is plotted in Fig. \ref{fig:COMP_MSR_T}  and shows a similar behavior as the measured noise PSD. In first region of the curves (sampling frequency up to $\lambda'_H$), the noise decreases with increase in the sampling frequencies. In this region, the stationary noise part of Eq. (\ref{eq:Sformula_alter}) dominates over the switching noise PSD. In the remaining portion (sampling frequencies higher than $\lambda'_H$), the noise is from the switched noise PSD as given in Eq. (\ref{eq:Sformula}) and first part of the Eq. (\ref{eq:Sformula_alter}).
 \par
In the noise simulation (PSS + Pnoise analysis) using Spectre, the noise reduction does not show any variation for different switching frequencies. However in measurements it was observed that as the switching frequency changes from 1 KHz to 1 MHz (for 6 stages), an additional noise reduction of 3.7 dB is obtained. Thus, the noise model used in Spectre is inconsistent with both the theoretical model and experimental results. The exact cause of this inconsistency of the proposed model and measured results requires more analysis.

\appendix

\section{Time domain autocovariance analysis to derive the RTS noise PSD in a variable duty cycle switched biasing condition}
As the trap state $N(t)$, for periodic biasing conditions, is considered as a cyclo-stationary process, here time averaged ACF function, for $N(t)$, is derived for one time period. A pulsed wave with time period $T$ with ON time of $T/n$, is applied to the gate of MOS transistor as the biasing signal. Let's suppose $p_{\text{\textit{on}}}(t)$ is the PTO at time $t$ during ON time of the device. As $p_{\text{\textit{on}}}(t)$ is a time varying function, the probability of any event in very small time interval $\Delta t$, can be given as:
\begin{equation}
\begin{aligned}
p_{\text{\textit{on}}}(t + \Delta t) = p_{\text{\textit{on}}}(t). \text{(probability of zero emission event in ~~~~ } \\ 
\text{$\Delta$t}) + (1-p_{\text{\textit{on}}}(t)). \text{(probability of acapture event in $\Delta$t}),~~~~~~ \nonumber 
\end{aligned}
\end{equation}
\begin{equation}
 ~~~~~~= p_{\text{\textit{on}}}(t).(1-\lambda_e \Delta t) + (1-p_{\text{\textit{on}}}(t))\lambda_c  \Delta t. \vspace{- 0.05 in}
\end{equation}
As for ON state of transistor $\lambda_c  \approx  \lambda_e   \approx
\lambda_{\text{\textit{on}}}$, the probability of a capture or emission event in $\Delta$t time is equal to $\lambda_{\text{\textit{on}}}$$\Delta$t and if $\Delta$t is infinitesimally small, (17) can be simplified as:
\begin{equation}
\frac{dp_{\text{\textit{on}}}(t)}{dt} + 2p_{\text{\textit{on}}}(t)\lambda_{\text{\textit{on}}} = \lambda_{on}. \vspace{- 0.05 in}
\end{equation}  
\begin{equation}
P_{\text{\textit{on}}}(t)= 0.5 + a e^{-2\lambda_{\text{\textit{on}}}t}, \hspace{0.3cm} 0 \leq t < T/n,
\end{equation}  
where `a' is the initial PTO during ON time.

Similarly $p_{\text{\textit{off}}}(t)$ is the PTO at time $t$, during OFF time of the device. As $p_{\text{\textit{off}}}(t)$ is also a time-varying function, the probability of any event in infinitesimally small time interval $\Delta t$, can be given as:\vspace{- 0.05 in}
\begin{equation}
\begin{aligned}
p_{\text{\textit{off}}}(t + \Delta t) = p_{\text{\textit{off}}}(t). \text{(probability of zero emission event}\\ 
 \text{in $\Delta$t}) + (1-p_{\text{\textit{off}}}(t)). \text{(probability of acapture event in $\Delta$t}), \nonumber
\end{aligned}
\end{equation}
\begin{equation}
p_{\text{\textit{off}}}(t + \Delta t) = p_{\text{\textit{off}}}(t).(1-\lambda_e \Delta t) + (1-p_{\text{\textit{off}}}(t))\lambda_c  \Delta t.  
\end{equation}
For OFF state of transistor, the probability of trap to capture an electron is almost zero, as the number of charge carriers are negligible, then $\lambda_c \approx 0$ and $\lambda_e = \lambda_{\text{\textit{off}}}$,  
\begin{equation}
p_{\text{\textit{off}}}(t + \Delta t) = p_{\text{\textit{off}}}(t).(1-\lambda_off \Delta t).  
\end{equation}
\begin{equation}
dp_{\text{\textit{off}}}(t)/dt = - p_{\text{\textit{off}}}(t)\lambda_{\text{\textit{off}}}.
\end{equation}  
\begin{equation}
P_{\text{\textit{off}}}(t)= b e^{-\lambda_{\text{\textit{off}}}t}, \hspace{0.5cm} \text{for} \hspace{0.3cm} T/n \leq t < T,
\label{eq:isub1}
\end{equation}
where `$b$' is the constant depends on the PTO at initial condition of the OFF state of transistor. The value of `$a$' and `$b$' can be arrived as:
\begin{equation}
\begin{aligned}
a = -(1-\beta)/(1-\alpha\beta); \hspace{0.3cm}  b = (1-\alpha)/(1-\alpha\beta),
\label{eq:isub2}
\end{aligned}
\end{equation}
\begin{equation}
\text{where} ~~~\alpha = e^{ -2\lambda_{\text{\textit{on}}} \frac{T}{n} }, \hspace{0.3cm} \beta = e^{ -\lambda_{\text{\textit{off}}} \frac{(n-1)T}{n} }. ~~~~~~~~~~~ \nonumber
\end{equation}
In order to calculate the ACF function for trap state $N(t)$, it is necessary to calculate condition probability of trap to be filled at time $(t+\tau)$, with the condition that $p(t) = 1$ (applicable for both ON as well as OFF time). This way conditional probabilities for trap occupancy at some arbitrary time $(t+\tau)$, with the condition that trap is full at time $t$, is:  
\begin{equation}
P_{on,11}(t)= \frac{1}{2}(1 + e^{-2\lambda_{\text{\textit{on}}}|\tau|}), \hspace{0.3cm} 0 \leq t < T/n,\hspace{0.2cm} t \geq 0.
\end{equation}  
\begin{equation}
P_{\text{\textit{off,11}}}(t)= e^{-\lambda_{\text{\textit{off}}}|\tau|}), \hspace{0.3cm} 0 \leq t < (T-1)/n,\hspace{0.2cm} t \geq 0.
\vspace{- 0.2 in}
\end{equation}  \\

Derivation of time averaged ACF function $C_{\lambda}(t,\tau)$ of $N(t)$ for ON and OFF time of transistor:
\begin{equation}
\begin{aligned}
C_{\lambda}(t,\tau) = \mathbb{E} [N(t - \tau/2).N(t + \tau/2)] \hspace{1.8cm}\\
\hspace{1 cm} - \mathbb{E}[N(t - \tau/2)].\mathbb{E}[N(t + \tau/2)] .
\end{aligned}
\end{equation}
ACF function $C^\tau_{\lambda,on}(t,\tau)$ for ON time 
\begin{equation}
\begin{aligned}
C_{\lambda,on}(t,\tau) = p_{\text{\textit{on}}}(t - \tau/2)p_{on,11}(\tau) \hspace{2.5cm} \\
 -~ p_{\text{\textit{on}}}(t - \tau/2)p_{on,11}(t + \tau/2), \nonumber 
\end{aligned}
\end{equation}
\begin{equation}
\begin{aligned}
= 0.25 e^{-2\lambda_{\text{\textit{on}}}\left|\tau\right|} - a^2 e^{-4\lambda_{\text{\textit{on}}}t}, ~\text{for}~\left|\tau/2\right| \leq t < T/n - \left|\tau/2\right| \\
\text{with}~ \left|\tau \right| \leq T/n. ~~~~~~~~~~~
\end{aligned}
\end{equation}
Time average of $C^\tau_{\lambda,on}(t,\tau)$ for ON state of the transistor can be given as:
\begin{equation}
\begin{aligned}
C^s_{\lambda,on}(\tau) = \frac{1}{T} \int_{\left|\frac{\tau}{2}\right|}^{\frac{T}{n} - \left|\frac{\tau}{2}\right|} C^\tau_{\lambda,on}(t,\tau)dt, \hspace{2.4cm} ~~ \nonumber
\end{aligned}
\end{equation}
\begin{equation}
\begin{aligned}
= (1/4T) [e^{-2\lambda_{\text{\textit{on}}}\left|\tau\right|}(T/n-a^2/\lambda_{\text{\textit{on}}}-\left|\tau \right|)~~~~ \\
+~(a^2/\lambda_{\text{\textit{on}}})(e^{-\frac{4\lambda_{\text{\textit{on}}}T}{n}}.e^{2\lambda_{\text{\textit{on}}}\left|\tau \right|})],~~\text{for}~|\tau| \leq T/n.
\end{aligned}
\end{equation}
The Fourier transform of time averaged ACF gives the noise PSD. The RTS noise PSD for ON state of switched bias transistor, with a variable duty cycle, can be given as:
\begin{equation}
\begin{aligned}
S^{\text{\textit{on}}}_\lambda(\omega) = \mathcal{F}(C^s_{\lambda,on}(\tau)) =  \int_{-\frac{T}{2}}^{\frac{T}{2}} C^s_{\lambda,on}(\tau)e^{-j\omega\tau}d\tau. 
\end{aligned}
\end{equation}
\begin{equation}
\begin{aligned}
S_{\lambda,{\text{\textit{on}}}}(\omega) =
\frac{1}{4T(4\lambda^2_{\text{\textit{on}}}+\omega^2)} \Biggl[ \frac{4T\lambda_{\text{\textit{on}}}}{n}-A -\frac{Be^{\frac{-2\lambda_{\text{\textit{on}}}T}{n}}}{4 \lambda^2_{on}+\omega^2} \Biggr],
\label{eq:isub6} ~~~~~~~\\
\\
A = a^2 4e^{\frac{-4\lambda_{\text{\textit{on}}}T}{n}} + \frac{ (16a^2 + 8)\lambda^2_{\text{\textit{on}}} + a^24\omega^2 -2\omega^2}{(4\lambda^2_{\text{\textit{on}}}+\omega^2)},~~~~~~~~~~~~~~ \\
\\
B = (40a^2 + 6) \lambda^2_{\text{\textit{on}}}cos(\omega T/n) - 8\omega \lambda_{\text{\textit{on}}}sin(\omega T/n). ~~~~~~~~~~~~~~
\end{aligned}
\end{equation}

\begin{acknowledgments}
The authors thank Space Applications Center (SAC)$-$ISRO,~Ahmedabad,~India for supporting in this work. 
\end{acknowledgments}


\end{document}